\documentclass[a4,12pt]{extreport}
\usepackage[sfdefault]{FiraSans}

\usepackage[utf8]{inputenc}
\usepackage[T1]{fontenc}

\usepackage{titlesec}
\usepackage{tocloft}
\usepackage[table]{xcolor}
\usepackage[export]{adjustbox}
\usepackage{wrapfig}
\usepackage{lipsum}
\usepackage{soul}
\usepackage[backgroundcolor=black!10,bordercolor=white!70!black,linecolor=white!70!black]{todonotes}
\usepackage{breakcites}
\usepackage{varioref}
\usepackage{amsfonts}
\usepackage{amssymb}
\usepackage{pifont}
\usepackage{mathtools}
\usepackage{oplotsymbl}
\usepackage{array}
\usepackage{xifthen}
\usepackage{xhfill}
\usepackage{geometry}
\usepackage[graphicx]{realboxes}
\usepackage{booktabs}
\usepackage{makecell}
\usepackage{multicol}
\usepackage{multirow}
\usepackage{colortbl}
\usepackage{varwidth}
\usepackage{setspace}
\usepackage{tabularx}
\usepackage{longtable}
\usepackage{arydshln}
\usepackage{pdfpages}
\usepackage[lined]{algorithm2e} \usepackage{semantic}
\usepackage{fancyhdr}
\usepackage{lastpage}
\usepackage[document]{ragged2e}
\usepackage{hyphenat}
\usepackage[most]{tcolorbox}
\usepackage{caption}
\usetikzlibrary{backgrounds}

\newcommand{\insertauthor}{Roberto Metere, Myriam Neaimeh, Charles Morisset, Carsten Maple, Xavier Bellekens, Ricardo M. Czekster}
\newcommand{\titleA}{Securing the}
\newcommand{\titleB}{Electric Vehicle}
\newcommand{\titleC}{Charging Infrastructure}
\newcommand{\inserttitle}{\titleA\ \titleB\ \titleC}
\newcommand{\insertsubtitle}{State-of-the-art review and recommendations with a focus on smart charging and vehicle-to-grid}

\definecolor{palette15}{HTML}{312f38}
\definecolor{palette25}{HTML}{545066}
\definecolor{palette35}{HTML}{9a8aa6}
\definecolor{palette45}{HTML}{e3c5d5}
\definecolor{palette55}{HTML}{e4d39f}

\colorlet{text}{palette15}
\colorlet{titles}{palette25}
\colorlet{subtitles}{palette35}
\colorlet{subsubtitles}{palette45}

\PassOptionsToPackage{hyphens}{url}\usepackage{hyperref} \hypersetup{
    pdfauthor={\insertauthor}
  , pdfkeywords={power system,uk}
  , pdftitle={\inserttitle}
  , colorlinks , linkcolor=
  , linkbordercolor={palette55}
  , pdfborderstyle={/S/U/W 1}
  , citecolor={palette35}
  , urlcolor={palette35}
  , bookmarksopen
  , bookmarksdepth=2
}
\usepackage{hypcap}
\usepackage{glossaries}
\usepackage[backend=bibtex]{biblatex}
\bibliography{bibliography}

\usepackage{framed}

\newif\ifxetexorluatex
\ifxetex
  \xetexorluatextrue
\else
  \ifluatex
    \xetexorluatextrue
  \else
    \xetexorluatexfalse
  \fi
\fi
\ifxetexorluatex \usepackage{fontspec}
  \usepackage{libertine} \newfontfamily\quotefont[Ligatures=TeX]{Linux Libertine O} \else
  \usepackage[utf8]{inputenc}
  \usepackage[T1]{fontenc}
  \usepackage{libertine} \newcommand*\quotefont{\fontfamily{LinuxLibertineT-LF}} \fi

\newcommand*\quotesize{60} \newcommand*{\openquote}
   {\tikz[remember picture,overlay,xshift=-4ex,yshift=-2.5ex]
   \node (OQ) {\quotefont\fontsize{\quotesize}{\quotesize}\selectfont``};\kern0pt}

\newcommand*{\closequote}[1]
  {\tikz[remember picture,overlay,xshift=4ex,yshift={#1}]
   \node (CQ) {\quotefont\fontsize{\quotesize}{\quotesize}\selectfont''};}

\colorlet{shadecolor}{white!98!black}

\newcommand*\shadedauthorformat{\emph} 

\newcommand*\authoralign[1]{\if#1l
    \def\authorfill{}\def\quotefill{\hfill}
  \else
    \if#1r
      \def\authorfill{\hfill}\def\quotefill{}
    \else
      \if#1c
        \gdef\authorfill{\hfill}\def\quotefill{\hfill}
      \else\typeout{Invalid option}
      \fi
    \fi
  \fi}
\newenvironment{shadequote}[2][r]{\authoralign{#1}
\ifblank{#2}
   {\def\shadequoteauthor{}\def\yshift{-2ex}\def\quotefill{\hfill}}
   {\def\shadequoteauthor{\par\authorfill\shadedauthorformat{#2}}\def\yshift{2ex}}
\begin{snugshade}\begin{quote}\openquote}
{\shadequoteauthor\quotefill\closequote{\yshift}\end{quote}\end{snugshade}}

\newcommand{\sidenote}[1]{\todo[size=\scriptsize,color=black!20!blue!10!white,bordercolor=black!20]{#1}}
\newcommand{\sn}[2][]{{\ifthenelse{\isempty{#1}}{\sidenote{#2}}{\sidenote{{\bf #1:} #2}}}}
\newcommand{\csn}[3][]{{\setulcolor{red}\ul{#3}}{\ifthenelse{\isempty{#1}}{\sidenote{#2}}{\sidenote{{\bf #1:} #2}}}}

\newcommand{\bulletnote}[3][\ding{202}]{{\hspace{0.3em}\color{#3}\small\raisebox{#2}{#1}}}
\definecolor{ncl}{rgb}{0.86,0.06,0.18}
\definecolor{ati}{rgb}{0,0,0}
\definecolor{uow}{rgb}{0.1,0.36,0.68}
\definecolor{uos}{rgb}{0.78,0.65,0}
\newcommand{\ncl}[1][0.3em]{\bulletnote[\ding{192}]{#1}{ncl}}
\newcommand{\ati}[1][0.3em]{\bulletnote[\ding{193}]{#1}{ati}}
\newcommand{\uow}[1][0.3em]{\bulletnote[\ding{194}]{#1}{uow}}
\newcommand{\uos}[1][0.3em]{\bulletnote[\ding{195}]{#1}{uos}}

\newcommand{\hr}{\hbox to\headwidth{\color{palette55}\leaders\hrule height \headrulewidth\hfill}}
\newcommand{\uld}[1]{\parbox[b][-1pt][l]{0pt}{\color{palette55}\leavevmode\makebox[\widthof{#1\,}]{\xleaders\hbox{.}\hfill\kern0pt}}#1}
\newcommand{\well}[2][]{\bgroup\small\noindent \renewcommand{\arraystretch}{1.5}
\begin{tabular*}{\textwidth}{m{0.00001\textwidth}m{0.013\textwidth}m{0.87\textwidth}p{0.001\textwidth}}& \cellcolor{palette45}\color{white}\centering\rotatebox{90}{\textsc{\bfseries #1}}& {\color{gray}#2}& \end{tabular*}\egroup}

\titlespacing\subsubsection{0pt}{1em}{-0.1em}

\hbadness = \tolerance
\makeatletter
\patchcmd{\@makechapterhead}{#1}{\hyphenpenalty=10000 #1}{}{}\patchcmd{\@makeschapterhead}{#1}{\hyphenpenalty=10000 #1}{}{}\makeatother
\newcommand{\chsep}{\hspace{20pt}}
\newcommand{\chrule}{{\color{palette55}\vrule width 3pt}}
\titleformat{\chapter}[hang]{\color{titles}\huge\bfseries}{\thechapter\chsep\chrule\chsep}{0pt}{}
\titleformat{\section}[hang]{\color{subtitles}\Large\bfseries}{\thesection\chsep}{0pt}{}
\titleformat{\subsection}[hang]{\color{subtitles}\large\bfseries}{\thesubsection\chsep}{0pt}{}
\titleformat{\subsubsection}[hang]{\color{subsubtitles}\normalsize\bfseries}{}{0pt}{}
\newcommand{\tnl}{\\\hspace{2.1cm}}
\newcommand{\stnl}{\\\hspace{1.5cm}}

\newcommand{\pagecounter}{\vspace{0.1cm}\ooalign{\color{text} \rule{3cm}{1.9cm} \cr
  \hfil\strut\raisebox{0.6cm}{\sffamily\bfseries\color{palette55}\Huge \thepage
  \kern0.1em\rule{1pt}{1.4ex}\kern0.15em\Large \pageref*{LastPage}}\hfil
}}
\fancypagestyle{e4future}{\fancyhf{}
  \renewcommand{\headrulewidth}{2pt}
  \fancyhead[L,R]{}
  \fancyhead[C,C]{\color{titles}\bfseries\textsc{\leftmark}}
  \fancyhead[R,L]{}
  
  \fancyfoot[R,L]{\vspace{2em}\color{titles}\sc{\inserttitle}}
  \fancyfoot[C,C]{}
  \fancyfoot[R]{\pagecounter}
}
\fancypagestyle{plain}{\fancyhf{}
  \renewcommand{\headrulewidth}{2pt}
  \fancyhead[L,R]{}
  \fancyhead[C,C]{}
  \fancyhead[R,L]{}
  
  \fancyfoot[R,L]{\vspace{2em}\color{titles}\sc{\inserttitle}}
  \fancyfoot[C,C]{}
  \fancyfoot[R]{\pagecounter}
}

\colorlet{evenrows}{white}
\colorlet{oddrows}{palette55!20!white}
\newcommand{\logot}[3][\textwith]{\begin{minipage}{#1}\centering\includegraphics[width=\textwidth,max height=#2]{#3}\end{minipage}}

\newcommand{\thr}[2][\rotation]{\rotatebox{#1}{\bfseries\color{titles}#2}}
\newcommand{\thh}[1]{\bfseries\color{titles}#1}
\newcommand{\thc}[2][\centering]{\cellcolor{oddrows}#1#2}
\newcommand{\continuetable}[1]{\multicolumn{#1}{r}{\cellcolor{white}\color{subtitles}{\small \em continued on next page}}}
\newcommand{\continuedtable}[1]{\multicolumn{#1}{r}{\cellcolor{white}\color{subtitles}{\small \em continued from previous page}}}

\newcommand{\rb}{\footnotesize\circletfill}

\newcommand{\icon}[2][inline]{\ifthenelse{\equal{#1}{inline}}{\hspace{1.4em}\begin{tikzpicture}[remember picture,overlay]
  \node[] at (-0.9em,0.4em) { \includegraphics[width=1.3em]{#2} };
\end{tikzpicture}}{\begin{tikzpicture}[remember picture,overlay]
  \node[] at (#1) { \includegraphics[width=1.3em]{#2} };
\end{tikzpicture}}}

\newcommand{\icongui}[1][inline]{\icon[#1]{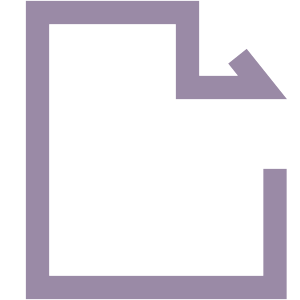}}
\newcommand{\iconrec}[1][inline]{\icon[#1]{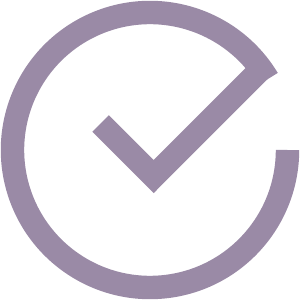}}

\newcommand{\tab}[3][2]{scritta-icona}

\newcommand{\reconly}{\iconrec[0em,1em]\newline}
\newcommand{\guionly}{\icongui[0em,1em]\newline}
\newcommand{\guirec}{\icongui[-0.7em,1em]\iconrec[0.9em,1em]\newline}

\newtcbox{\pill}[1][palette55]{nobeforeafter,tcbox raise base,arc=0.6em,outer arc=0.6em,top=0.1em,bottom=0.1em,left=0.2em,right=0.2em,leftrule=0mm,rightrule=0mm,toprule=0mm,bottomrule=0mm,boxsep=0.1em,colback=#1,colframe=#1,coltext=white,fontupper=\bfseries}
\newcommand{\refpill}[2][palette55]{\pill[#1]{§\ref{#2}}}

\newcolumntype{C}{>{\centering\arraybackslash\color{titles}\bfseries}m{0.15\textwidth}} \newcolumntype{D}{>{\raggedright\arraybackslash\color{text}}m{0.16\textwidth}} \newcolumntype{E}{>{\raggedright\arraybackslash\color{text}}m{0.36\textwidth}} \newcolumntype{M}{>{\centering\arraybackslash\color{text}}b{0.014\textwidth}} 

\newcolumntype{V}{>{\arraybackslash\color{titles}\bfseries}m{0.35\textwidth}} \newcolumntype{N}{>{\arraybackslash\color{text}}m{0.6\textwidth}} 

\newcolumntype{Y}{>{\centering\arraybackslash\vspace{-1em}\color{green!70!black}$\blacktriangleright$}p{0.014\textwidth}} \newcolumntype{S}{>{\arraybackslash\vspace{-.5em}\color{text}}p{0.9\textwidth}} \newcolumntype{L}{>{\centering\arraybackslash\vspace{-.5em}\color{green!70!black}}p{0.014\textwidth}} \newcolumntype{Z}{>{\centering\arraybackslash\color{green!70!black}\openup .3em}p{0.05\textwidth}} \newcolumntype{R}{>{\arraybackslash\vspace{-1em}\color{text}}p{0.95\textwidth}} 

\newcommand{\acr}[1]{\hypertarget{#1}{#1}}
\newcommand{\gl}[1]{\protect\hyperlink{#1}{\uld{#1}}}

\newcolumntype{T}{>{\raggedleft\arraybackslash\color{titles}\bfseries}p{0.22\textwidth}} \newcolumntype{F}{>{\raggedright\arraybackslash\color{text}}p{0.78\textwidth}} 

\newcolumntype{U}{>{\raggedleft\arraybackslash\color{titles}\bfseries}p{0.2\textwidth}} \newcolumntype{G}{>{\raggedright\arraybackslash\color{text}}p{0.8\textwidth}}

\counterwithout*{footnote}{chapter}

\begin{document}
\color{text}
\thispagestyle{empty}
\begin{tikzpicture}[remember picture,overlay]
  \node[] (picture) at (current page.center) { \includegraphics[width=\paperwidth]{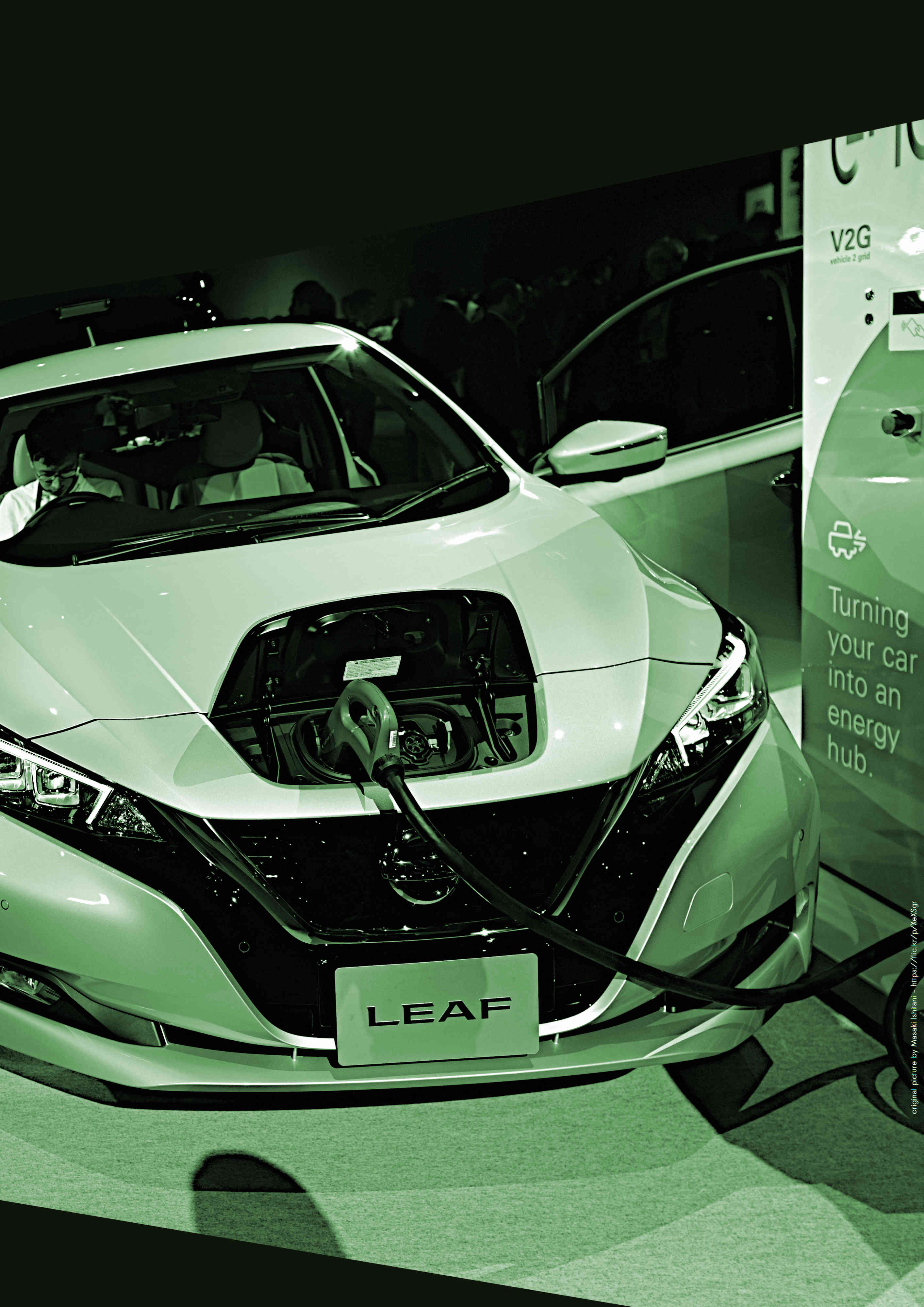} };
  \node[] (frontpage) at (current page.center) { \includegraphics[width=\paperwidth]{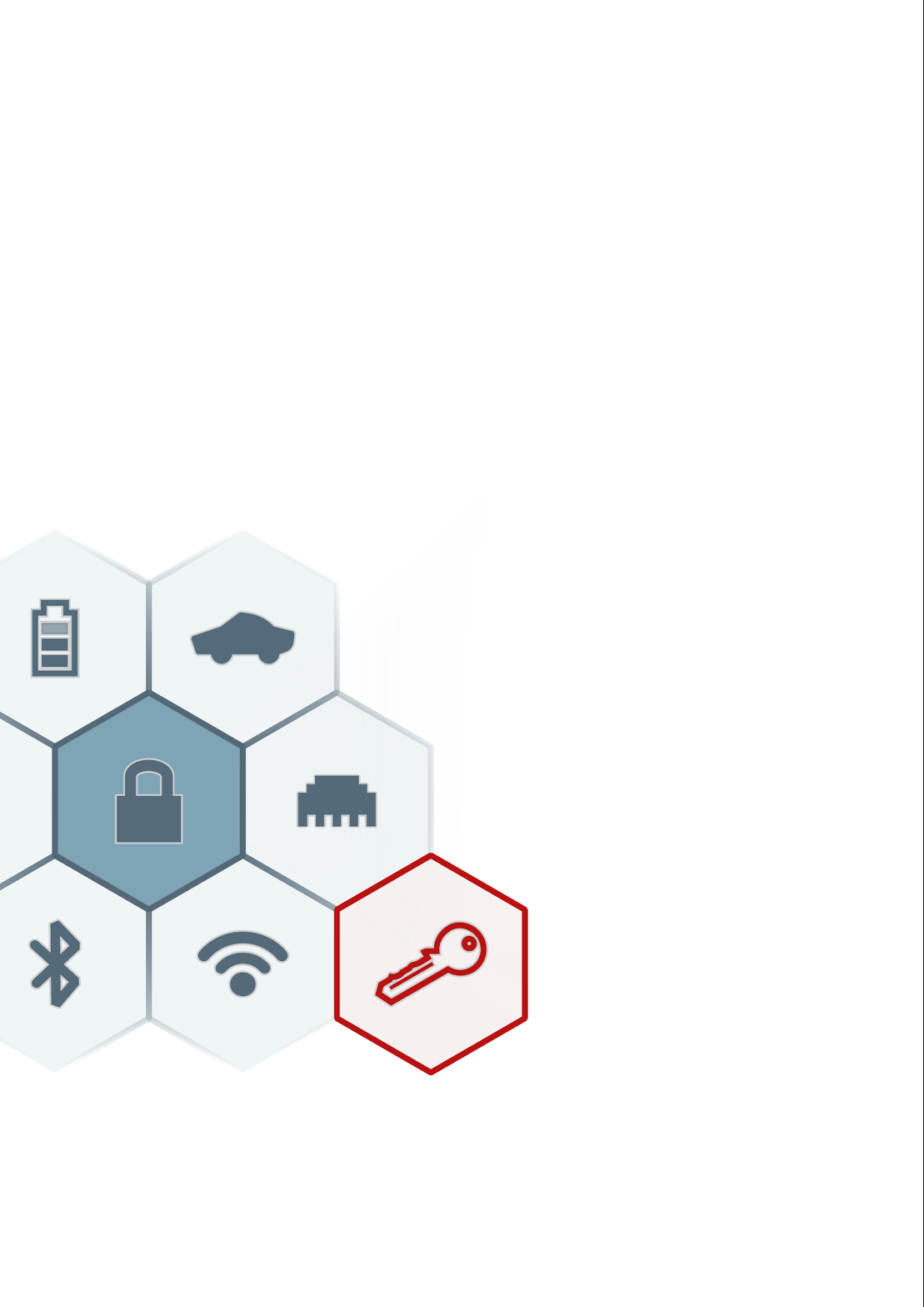} };
   
  \node[anchor=north west,xshift=3em,yshift=-4em] (names) at (current page.north west) {
    {\color{titles}\Huge\bfseries {\textsc{\titleA}}}
  };
  \node[anchor=north west,xshift=3em,yshift=-6.5em] (names) at (current page.north west) {
    {\color{titles}\Huge\bfseries {\textsc{\titleB}}}
  };
  \node[anchor=north west,xshift=3em,yshift=-9em] (names) at (current page.north west) {
    {\color{titles}\Huge\bfseries {\textsc{\titleC}}}
  };
  \node[anchor=north west,xshift=3em,yshift=-13em] (names) at (current page.north west) {
    \begin{varwidth}{0.5\paperwidth}
      {\color{subtitles}\Large\bfseries \nohyphens{\textsc{\insertsubtitle}}}
    \end{varwidth}
  };
  
  \node[anchor=north west,minimum height=4cm] (names) at ([xshift=1em,yshift=4.2cm]current page.south west) {
    \begin{tabularx}{0.7\paperwidth}{cccc}
        \includegraphics[height=1.2cm,valign=b]{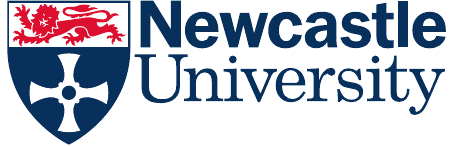}
      & \includegraphics[height=1.2cm,valign=b]{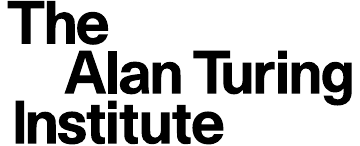}
      & \includegraphics[height=1.0cm,valign=b]{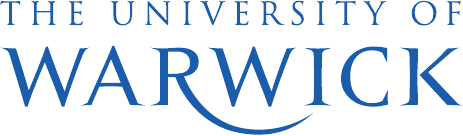}
      & \includegraphics[height=1.8cm,valign=b]{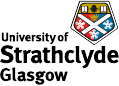}
    \end{tabularx}
    };

  \node[anchor=north east] (names) at ([xshift=-1em,yshift=2.4em]current page.south east) { \color{palette55!70!black}\bfseries April 2021 };
\end{tikzpicture}
\pagestyle{e4future}
\newpage
\bgroup \small
\section*{Authors}

\begin{minipage}[t]{0.49\textwidth}
 \subsubsection*{Roberto Metere\ncl\ati}
 is a Research Associate in Security in the School in Computing at Newcastle University. His work is supported by the e4Future grant and a grant from Turing's Data Centric Engineering programme, sponsored by the Lloyd's Register Foundation.
\end{minipage}\begin{minipage}[t]{0.02\textwidth}\strut\end{minipage}\begin{minipage}[t]{0.49\textwidth}
 \subsubsection*{Myriam Neaimeh\ncl\ati}
 is a Turing Research Fellow and Data-Centric Engineering Group Leader on Vehicle Grid Integration at the Alan Turing Institute. Her work is supported by the e4future grant and a grant from Turing's Data Centric Engineering programme, sponsored by the Lloyd's Register Foundation.
\end{minipage}
\begin{minipage}[t]{0.49\textwidth}
 \subsubsection*{Charles Morisset\ncl}
 is a Senior Lecturer in Security in the School in Computing Science at Newcastle University.
\end{minipage}\begin{minipage}[t]{0.02\textwidth}\strut\end{minipage}\begin{minipage}[t]{0.49\textwidth}
 \subsubsection*{Ricardo M. Czekster\ncl}
 is a Research Associate on the Active Building Centre Research Programme (ABC-RP) at Newcastle University.
\end{minipage}
\begin{minipage}[t]{0.49\textwidth}
 \subsubsection*{Xavier Bellekens\uos}
 is a Lecturer chancellor’s fellow in the Department of Electronic and Electrical Engineering at the University of Strathclyde and a Nonresident Fellow of the Scowcroft Center for Strategy and Security at the Atlantic Council.
\end{minipage}\begin{minipage}[t]{0.02\textwidth}\strut\end{minipage}\begin{minipage}[t]{0.49\textwidth}
 \subsubsection*{Carsten Maple\uow\ati}
 Professor Carsten Maple is the Principal Investigator of the NCSC-EPSRC Academic Centre of Excellence in Cyber Security Research at the University of Warwick and Professor of Cyber Systems Engineering in WMG.
 He is also a Fellow of the Alan Turing Institute, the National Institute for Data Science and AI in the UK where he is principal investigator of a \$5 million project developing trustworthy national identity.
 Carsten is co-investigator of PETRAS, the National Centre of Excellence for IoT Systems Cyber Security where he is the Sector Lead for Transport and Mobility.
 
\end{minipage}

\begin{center}
  \begin{tabularx}{0.7\paperwidth}{cccc}
      \ncl[0cm]
    & \ati[0cm]
    & \uow[0cm]
    & \uos[0cm]
    \\
      \includegraphics[height=1.2cm,valign=t]{res/ncl-logo.pdf}
    & \includegraphics[height=1.2cm,valign=t]{res/ati-logo.pdf}
    & \includegraphics[height=1.0cm,valign=t]{res/warwick.pdf}
    & \includegraphics[height=1.8cm,valign=t]{res/strath.pdf}
  \end{tabularx}
\end{center}

\begin{quotation}
\scriptsize\em\centering\noindent\color{gray}
This report has been prepared by a collaboration of {\bf Newcastle University}, {\bf The Alan Turing Institute}, {\bf University of Warwick}, and {\bf University of Strathclyde} and relates to the projects {\bf e4Future} and {\bf ABC-RP}.

The content and recommendations presented in this manuscript reflect the views of the authors alone.
\end{quotation}

\subsection*{e4Future}
e4Future is a real world demonstrator of Vehicle-to-Grid (V2G) charging technologies aiming to show how electric vans and cars can support the UK grid and provide a profitable and sustainable solution for business fleets.
The project is funded under UK Research and Innovation by the Department Department for Business, Energy and Industrial Strategy (BEIS) and The Office for Zero Emission Vehicles (OZEV) and delivered through Innovate UK.
e4Future is led by Nissan, in collaboration with energy companies and academic institutions.
Newcastle University leads the security work package on the project and a task to understand customer attitudes towards \gl{V2G}.
In addition, Newcastle University is building and simulating electricity distribution network models to understand impact and benefits of \gl{V2G}; and building a secure cloud based tool to automate collection and analysis of data coming from \gl{BEV}s, \gl{V2G} chargers and electricity networks.

Project partners are:

 \begin{tabular}{ccccccc}
    \logot[0.12\textwidth]{1.5cm}{res/ncl-logo.pdf}
  & \logot[0.12\textwidth]{1.5cm}{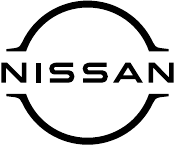}
  & \logot[0.12\textwidth]{1.5cm}{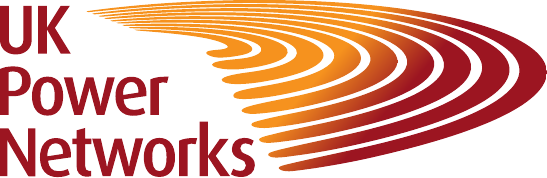}
  & \logot[0.12\textwidth]{1.5cm}{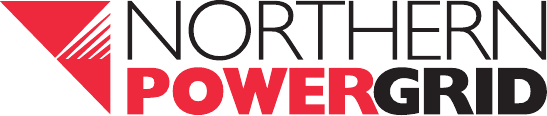}
  & \logot[0.12\textwidth]{1.5cm}{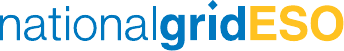}
  & \logot[0.12\textwidth]{1.5cm}{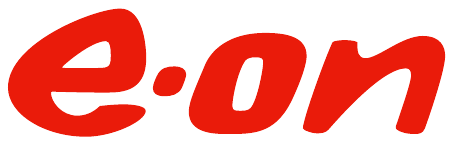}
  & \logot[0.12\textwidth]{1.5cm}{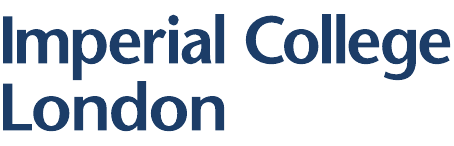}
 \end{tabular}

\subsection*{ABC-RP}
\begin{wrapfigure}{r}{0.4\textwidth}
  \includegraphics[width=0.4\textwidth]{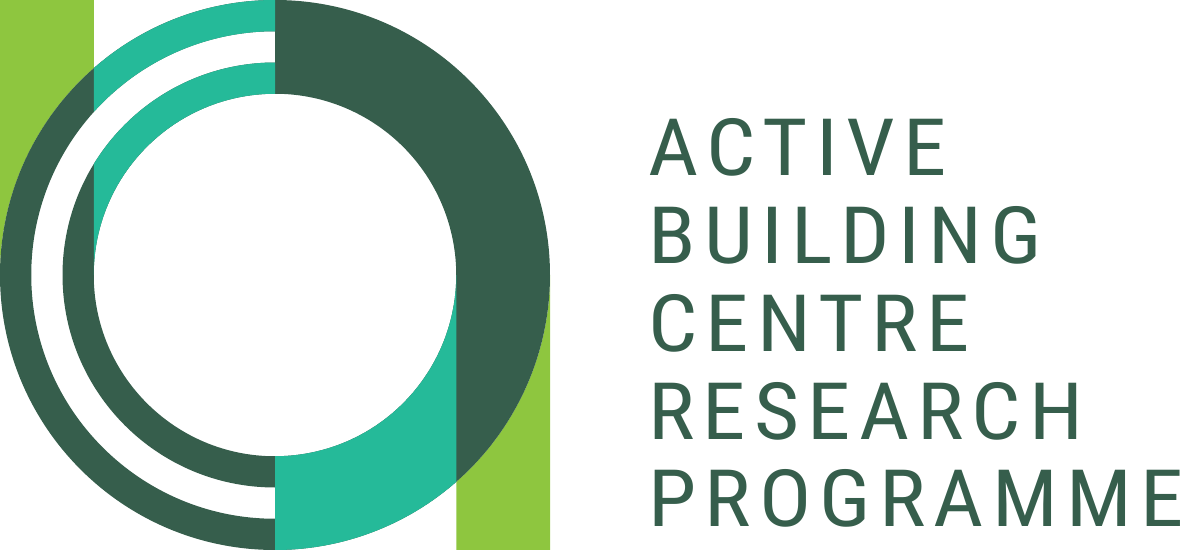}
\end{wrapfigure}
The Active Building Center Research Programme (ABC-RP) aims to address sustainable infrastructure to reduce the carbon footprint of buildings to meet governmental targets by 2050.
The consortium researches innovative ways and technologies to sustain high quality power provision to customers while regulating ancillary services in localised (small scale) grids.
The vision is to create ensembles of Active Buildings (ABs) acting as energy peers that trade the surplus.
They can energise their surroundings to adjust to flexible demands to meet dynamic energy requirements.
They are equipped with responsive technologies and state-of-the-art controls that combine near real-time decision making with remote management capabilities.

\gl{AB}s have the ability to operate disconnected from the conventional grid and in islanded mode (independent).
The infrastructure combines massive quantities of Distributed Energy Resources (DER), such as photo-voltaic, wind turbines, charge Energy Storage Systems (ESS) using large batteries attached to the buildings, or using pre-existing electric vehicles located in charging areas.

Link to the project: \url{https://abc-rp.com/}.

\egroup  
\section*{Acronyms}
\begin{minipage}[t]{0.44\textwidth}
  \begin{center}
    \rowcolors{2}{oddrows}{evenrows}
    \begin{longtable}{ T F }
    \multicolumn{2}{r}{\color{subtitles}{\small \em continued on next page}} \\ \endfoot
    \endlastfoot
      \acr{ACPO}     & Association of Chief Police Officers of England, Wales and Northern Ireland \\
      \acr{BEV}      & Battery Electric Vehicle \\
      \acr{CA}       & Certificate Authority \\
      \acr{CAN}      & Controller Area Network \\
      \acr{CAV}      & Connected and Autonomous Vehicles \\
      \acr{CCS}      & Combined Charging System \\
      \acr{CEC}      & China Electricity Council \\
      \acr{CPNI}     & Centre for the Protection of National Infrastructure \\
       \acr{CPO}     & Charge Point Operator \\
      \acr{DER}      & Distributed Energy Resources \\
      \acr{DfT}      & Department for Transport (UK) \\
      \acr{DoS}      & Denial of Service \\
      \acr{DSO}      & Distribution System Operator \\
      \acr{ECU}      & Electronic Control Units \\
      \acr{EMS}      & e-Mobility Services \\
      \acr{ENCS}     & European Network for Cyber Security \\
      \acr{EV}       & Electric Vehicle \\
      \acr{GB/T}     & (Recommended) Guobiao standards \\
      \acr{GDPR}     & General Data Protection Regulation \\
    \end{longtable}
  \end{center}
\end{minipage}\begin{minipage}{0.08\textwidth}\strut \end{minipage}\begin{minipage}[t]{0.44\textwidth}
  \begin{center}
    \rowcolors{2}{oddrows}{evenrows}
    \begin{longtable}{ T F }
    \multicolumn{2}{r}{\color{subtitles}{\small \em continued on next page}} \\ \endfoot
    \endlastfoot
    
      \acr{HEV}      & Hybrid Electric Vehicle \\
      \acr{ICO}      & Information Commissioner's Office \\
      \acr{IoT}      & Internet of Things \\
      \acr{ITS}      & Intelligent Transport System \\
      \acr{NIS}      & Network \& Information Systems \\
      \acr{NIST}     & National Institute of Standards and Technology \\
      \acr{OCPP}     & Open Charge Point Protocol \\
      \acr{PHEV}     & Plug-In Electric Vehicle \\
      \acr{PKI}      & Public-Key Infrastructure \\
      \acr{PnC}      & Plug\&Charge \\
      \acr{PUF}      & Physically Unclonable Functions \\
      \acr{SAE}      & Society of Automotive Engineers \\
      \acr{SIEM}     & Security Information and Event Management \\
      \acr{SoC}      & State of Charge \\
      \acr{TCU}      & Telematic Control Units \\
      \acr{TLS}      & Transport Layer Security \\
      \acr{TSO}      & Transmission System Operator \\
      \acr{V2G}      & Vehicle-to-Grid \\
    \end{longtable}
  \end{center}
\end{minipage}

\tableofcontents

\chapter{Executive Summary}

Electric Vehicles (EVs) can help alleviate our reliance on fossil fuels for transport and electricity systems.
However, charging millions of EV batteries requires management to prevent overloading the electricity grid and minimise costly upgrades that are ultimately paid for by consumers.

Managed chargers, such as Vehicle-to-Grid (V2G) chargers, allow control over the time, speed and direction of charging.
Such control assists in balancing electricity supply and demand across a green electricity system and could reduce costs for consumers.

Smart and V2G chargers connect EVs to the power grid using a charging device which includes a data connection to exchange information and control commands between various entities in the EV ecosystem.
This introduces data privacy concerns and is a potential target for cyber-security attacks.
Examples of threats include unauthorised access to
information (e.g. banking details),
tampering (e.g energy used), and
denial of service (e.g. unavailability of the charger) among others.
Therefore, the implementation of a secure system is crucial to permit both consumers and electricity system operators to trust smart charging and V2G.
 
In principle, we already have the technology needed for a connected EV charging infrastructure to be securely enabled, borrowing best practices from the Internet and industrial control systems.
The main tools would be cryptographic algorithms (e.g. AES-256) and protocols (e.g. TLS), digital signatures and certificates.
To deploy cryptographic systems on a widespread basis across multiple administrative domains, more work is required on governance and operational procedures such as defining a trust framework, for example a Public-Key Infrastructure, which is widely adopted on the Internet to securely exchange confidential messages and authenticate relevant entities.

We must properly adapt the security technology to take into account the challenges peculiar to the EV charging infrastructure.
Challenges go beyond technical considerations and other issues arise such as  balancing trade-offs between security and other desirable qualities such as interoperability, scalability, crypto-agility, affordability and energy efficiency.

This document reviews security and privacy topics relevant to the EV charging ecosystem with a focus on smart charging and V2G. 
\section{Recommendations and Guidance}
\label{sec:recommendations}

It is crucial to establish and adopt security requirements for the EV charging infrastructure before mass uptake of EVs.
Ensuring that an EV charging infrastructure, including smart charging and V2G, is secure requires security measures and standards at multiple levels from the charge point and electric vehicle through the charging infrastructure back office and grid operator.
Guidance and recommendations are provided to help secure charging infrastructure, electricity system and EV drivers.

\begin{center}
\renewcommand{\arraystretch}{1.3}
\setlength\arrayrulewidth{0.3em}
\arrayrulecolor{white}
\rowcolors{2}{palette45!20!white}{palette45!10!white}
\setlength{\leftmargini}{1em}
\renewcommand{\labelitemi}{\color{palette45!80!black}$\blacktriangleright$}
\begin{longtable*}{ZS}
  \multicolumn{2}{r}{\textsc{{\em Legend:\qquad} \bfseries \icongui Guidance \quad \iconrec Recommendation \quad \pill[palette55!90!black]{§ \textbullet} Section reference}} \\ \hline\hline
  \endfirsthead
  \continuedtable{2} \\ \endhead
  \continuetable{2} \\ \endfoot
  \endlastfoot
    \guionly
    \refpill[palette55!90!black]{ch:security-ev-charging-infrastructure}
    \refpill[palette55!90!black]{sec:privacy-regulations}
    &
  {\bf Regulations that apply to EV charging infrastructure have been introduced and some are under development.}
    Relevant players need to understand their roles and responsibilities in terms of existing applicable regulations such as {\bf GDPR} for data protection and the Security of Network \& Information Systems {\bf(NIS) regulations}~\cite{department2018nis}.
    Additionally, upcoming {\bf legislation on EV smart charging} ~\cite{gov2019evsc} in the UK will include cyber security requirements.
    Under the sponsorship of BEIS and OZEV, the British Standards Institution (BSI) is expected to publish in 2021 two publicly available specifications {\bf(PAS) 1878 and 1879} on energy smart appliances~\cite{bsi2020energy,klein2020uk}.
    These specifications include security requirements that, if adopted, can help secure smart and V2G charging infrastructure. \\ \hline
    \reconly
    \refpill[palette55!90!black]{ch:privacy-ev-charging-infrastructure}
    &
  {\bf Encryption of communications and the anonymisation of data should be mandatory to ensure privacy and protection of the EV ecosystem}.
  We envision that an enforcement authority such as the Office for Product Safety and Standards could collaborate with the Information Commissioner's Office (ICO) to provide data privacy guidance to relevant actors in the EV ecosystem and, when necessary, enforce fines for failure to comply with data protection laws.
    \\ \hline
    \guirec
    \refpill[palette55!90!black]{ch:protocols}
    &
  Several {\bf open communication protocols and standards} have been developed to manage distributed energy resources and smart appliances including EV chargers.
  Some are generic (e.g. OpenADR) while others are specifically designed for EVs and charging infrastructure (e.g. OCPP, ISO15118, among others).
  {\bf Most of these protocols include  provisions to meet cyber-security guidelines.}
    \\ \hline
    \reconly
    \refpill[palette55!90!black]{sec:nist-protect}
    &
  {\bf Publicly exposed devices should not allow unprotected physical access}, e.g. deploy tamper-proof chargers, {\bf and protect against remote attacks}, e.g. through authenticated and encrypted communications, including secure software and firmware updates.
    \\ \hline
    \reconly
    \refpill[palette55!90!black]{ch:security-ev-charging-infrastructure}
    \refpill[palette55!90!black]{sec:nist-protect}
    &
  {\bf Devices should be provided with appropriate memory and computational power to minimise insecure infrastructure or costly hardware updates}.
  Experience from other domains suggests that communication protocols are likely to change or be replaced and security keys could grow in size to offer stronger security guarantees.
    \\ \hline
    \guionly
    \refpill[palette55!90!black]{ch:security-ev-charging-infrastructure}
    &
  {\bf Technical documents are available to help develop secure charging infrastructure}, namely documents  published by the European Network for Cyber Security (ENCS) in collaboration with ElaadNL. These include a security test plan for EV charging stations and a proposed security architecture for charging infrastructure~\cite{elaadnl2019cybersecurity}.
    \\ \hline
    \guirec
    \refpill[palette55!90!black]{sec:nist-protect}
    &
  {\bf Security testing and assurance, as well as certifications and compliance tests are essential.} An example of a certification programme is the ENERGY STAR\textregistered\footnote{ENERGY STAR\textregistered is a US government-backed symbol that helps customers choose energy efficient products.} Program Requirements for Electric Vehicle Supply Equipment~\cite{energy2017energy}.
    \\ \hline
    \reconly
    \refpill[palette55!90!black]{ch:security-ev-charging-infrastructure}
    \refpill[palette55!90!black]{sec:nist-detect-respond-recover}
    &
  {\bf Security frameworks should be adopted and attack databases should be consulted to safeguard the EV ecosystem}.
  Organisations, e.g. point manufacturers, CPOs and aggregators, should carry out risk assessments and threat modellings to understand and manage security risks.
  Example of security framework is the NIST Cybersecurity Framework which provides organisations with guidance on how to prevent, detect, and respond to cyber attacks.
  Example of an attack database is MITRE's Adversarial Tactics, Techniques and Common Knowledge (ATT\&CK\textregistered).
    \\ \hline
    \guionly
    \refpill[palette55!90!black]{ch:protocols}
    &
  {\bf Alignment with international activities will be crucial.}
  A collaboration between industry, government and international partners is fundamental to ensure the UK is aligned with international markets.
  The publication of the {\bf European Cyber Security Act} has triggered the work on the {\bf Network Code on energy-specific cyber security} which is currently being developed~\cite{smart2017interim}.
  Network Codes have legal status within the European Union, and the UK has implemented previous codes.
  A large part of the European market adopting common standards and frameworks for EV charging cyber security would mean that adopting different approaches could potentially make the UK market less attractive to manufacturers and operators.
    \\ \hline
    \guirec
    \refpill[palette55!90!black]{sec:pki}
    &
  {\bf A platform to bring together emerging key players}, e.g. charge point manufacturers, aggregators, \dots, {\bf is recommended with the goal of sharing security issues, agreeing on requirements, developing capabilities and liaising with international activities}. Such platform could bring mobility and energy industries together to support the {\bf design of a PKI platform} for EV charging and {\bf identify options} to field such a platform in real world (e.g. one trust authority vs multi-regional trust authorities). It is worth noting that there are on-going European and international PKI platform design and development projects and it is not clear, to the best of our knowledge, if UK stakeholders are involved in these efforts.
  \\
\end{longtable*}
\end{center}

As in other contexts, the evolution of the security for the EV charging infrastructure is already a game between hackers and secure systems, with vulnerabilities and patches, mixed with compliance with local and international policies.
Thus, working tightly with security experts is fundamental to ensure that devices and processes put in place are properly secured.

 \chapter{Introduction}

Electric vehicles (EVs) can improve air quality and reduce our carbon emissions.
To support the adoption of EVs, several countries announced plans to end the sale of petrol and diesel cars (by 2030 in the UK)~\cite{pickett2021electric}.

With millions of EVs forecasted to be on the road in the coming years, charging their batteries needs to be managed to prevent overloading the electricity grid and minimise costly upgrades  that are ultimately paid for by consumers~\cite{neaimeh2015probabilistic}.

Unmanaged charging is still prevalent, where the battery of the EV will begin charging as soon as it is connected to the charging station until it is fully charged.

However, new managed charging technologies such as smart chargers and vehicle to grid (V2G) technologies are being introduced.
Electric vehicles, plugged using these new charging technologies, can help balance supply and demand across a green electricity system and reduce costs for the consumer~\cite{beis2020powering,flack2020cyber}.

Smart chargers and V2G connect an EV to the electricity grid using a charging device which includes a data connection to exchange information and control signals.
Both charging technologies would ensure that the vehicle is charged to meet driving requirements which the user can set, usually through a customer mobile application.

Smart chargers can manage the time of charging and the speed of charging (i.e. power rate).
For example, charging can start a couple of hours after the EV is plugged and it doesn't need to happen at full power.
Smart charging can minimise grid congestion by avoiding charging at peak times (7-9AM and 4-7PM in the UK) and/or make use of excess electricity generated by solar panels or grid electricity imported during cheaper periods.

A V2G charger is similar to a smart charger in that it can also manage the time and speed at which charging takes place.
In addition, a V2G charger allows switching the direction of charging by allowing EVs to sell energy back to the grid when it is most needed, or most expensive, such as at times of peak demand.

\section[An overview of the Smart Charging and Vehicle-to-Grid ecosystem]{\nohyphens{An overview of the Smart Charging and V2G ecosystem}}

Electric vehicles managed through Smart chargers or V2G chargers can provide services to the building/home (behind the meter services); the neighbourhood/local electricity networks (distribution system operator, DSO, services) and to the whole system/region (transmission system operator, TSO, services).
For more information on vehicle grid services, we refer to Andersen et al.~\cite{andersen2019parker}.
\begin{center}
  \includegraphics[width=.7\textwidth]{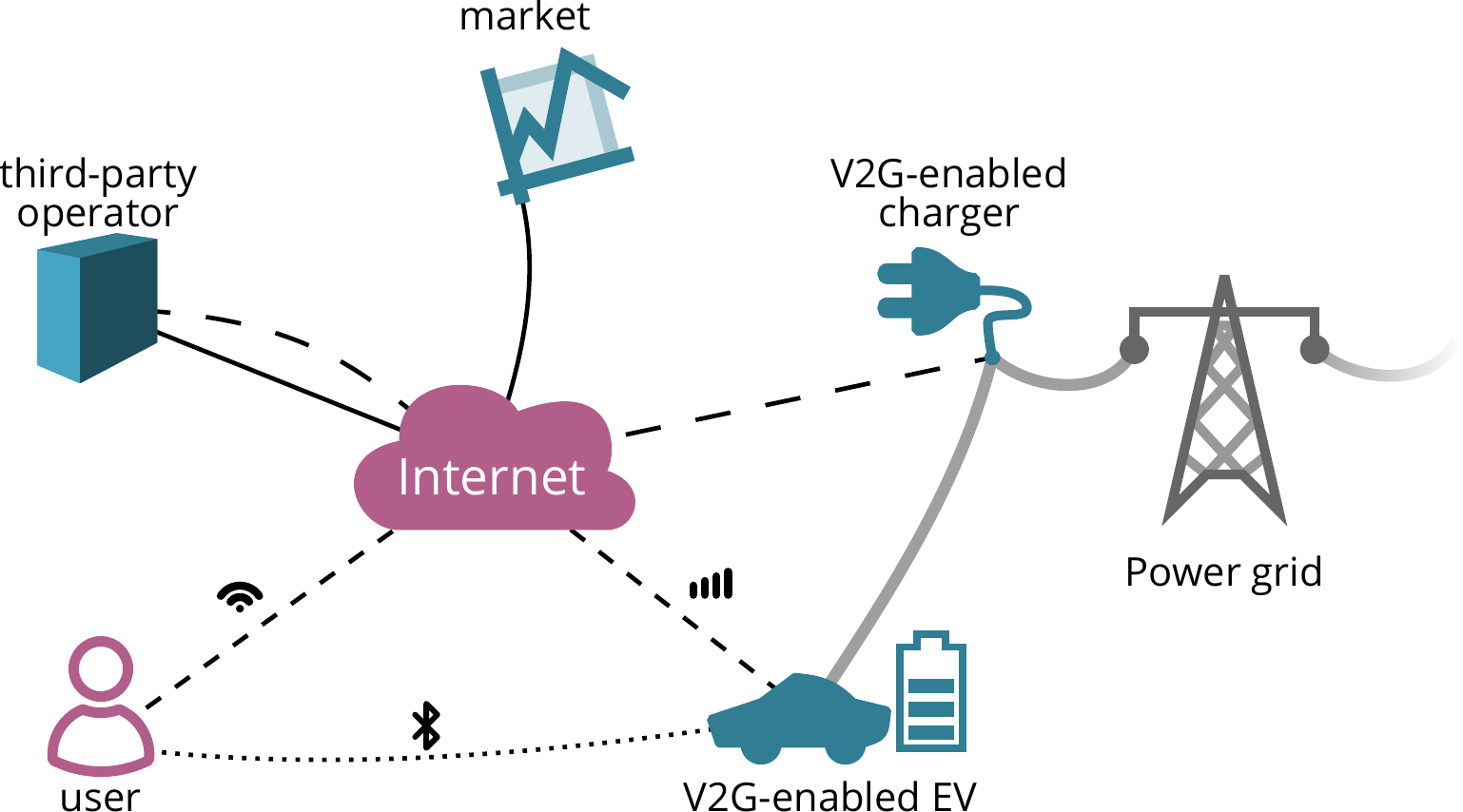}
\end{center}

\begin{figure}[htbp]
  \caption{Typical charging locations for private passenger BEVs -- adapted from Danish Electric Vehicle Alliance, DTU (2019) Sådan skaber Danmark grøn infrastruktur til én million elbiler.}
  \label{fig:v2g-charging-destinations}
  \begin{center}
    \includegraphics[width=.8\textwidth]{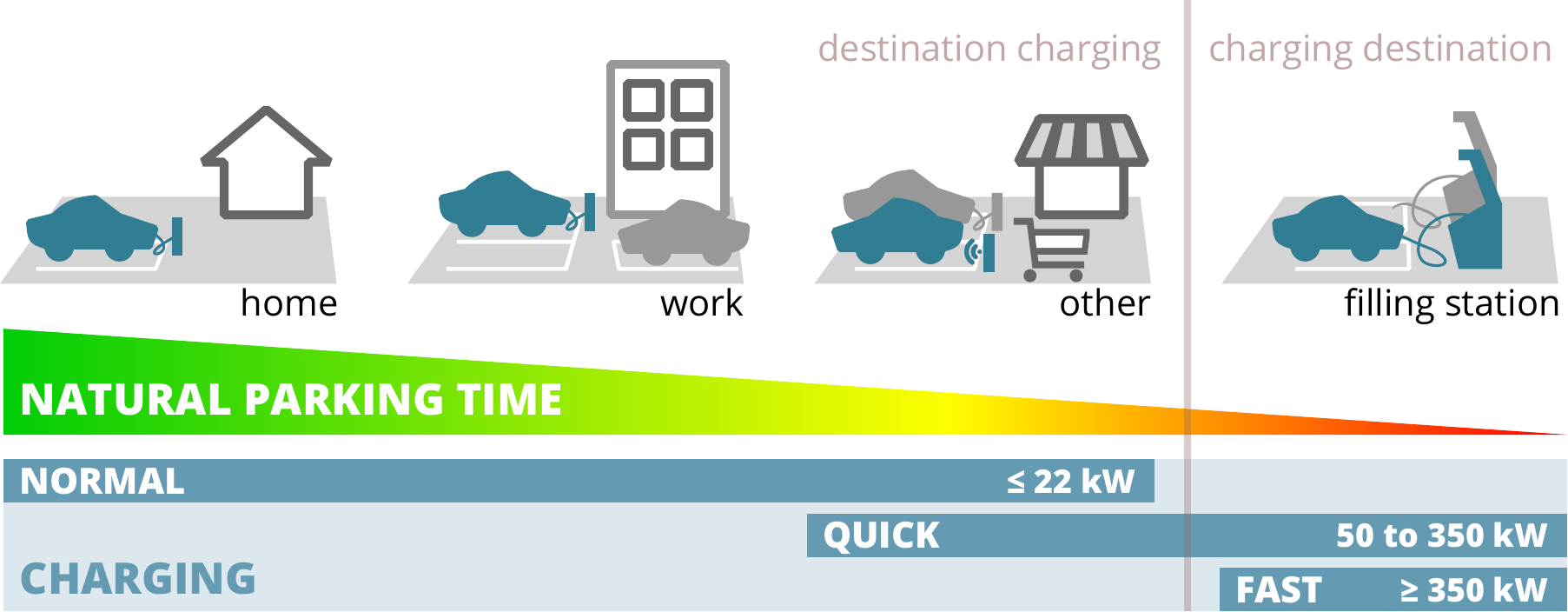}
  \end{center}
\end{figure}

Installing Smart chargers or V2G chargers makes sense at locations where vehicles are routinely parked for a long period of time such as residential and workplace locations, see Figure~\vref{fig:v2g-charging-destinations}. When cars are plugged for a long period of time, then there will be a flexibility to manage the time and speed of charge and in the case of V2G manage charging and discharging of the battery while prioritising the users' driving requirements.
\newpage
\section[Securing the Web of Energy: The importance of Privacy and Cybersecurity for charging stations]{\nohyphens{Securing the Web of Energy: The importance of Privacy \stnl and Cybersecurity for charging stations}}

Smart charging and vehicle-to-grid (V2G) charging technologies connect an EV to the electricity grid using a charging device which includes a data connection and ensures the exchange of information and control commands between various entities in the EV ecosystem including charge point, charge point operators, grid network operators, among others.

Any connected infrastructure is  a potential target for cyber-security attacks, with motives including information theft, cyber-warfare, or organised crime\footnote{
Examples are the attack against the power grid in Ukraine, during which 250,000 homes were cut off power for 6 hours~\cite{lee2016analysis}, and the WannaCry ransomware attack that paralysed critical infrastructure, such as the UK NHS, asking for ransom money to prevent the irreversible deletion of crucial files~\cite{martin2018wannacry}.}.

As such, we need to ensure the cyber security of these connected devices and the entities involved in operating them to prevent threats impacting both the consumers and the electricity system as a whole.

Moreover, privacy of users should be considered.
Sharing users' data by default without engaging the user is risky as they might resist participating in managed charging initiatives.
This is specifically relevant to users who do not trust ``smart'' systems and could resist using them. As an example, using charging points could give away user location, a perennial challenge with services depending on users' location.
Consequently, an EV charging ecosystem should consider techniques and processes to ensure user anonymity.

 \chapter[Security of the EV Charging Infrastructure]{Security of the EV Charging \tnl Infrastructure}
\label{ch:security-ev-charging-infrastructure}

The US National Institute of Standards and Technologies (NIST) Cybersecurity Framework provides organisations with guidance on how to prevent, detect, and respond to cyber attacks. NIST has categorised high-level, strategic view of the lifecycle of an organization's management of cybersecurity risk into five cybersecurity functions that are not synchronised nor disjoint~\cite{nist2018cybersecurity,cpni2015passport,cis2018cybersecurity}: Identify, Protect, Detect, Respond, and Recover.
\begin{center}
  \includegraphics[width=\textwidth]{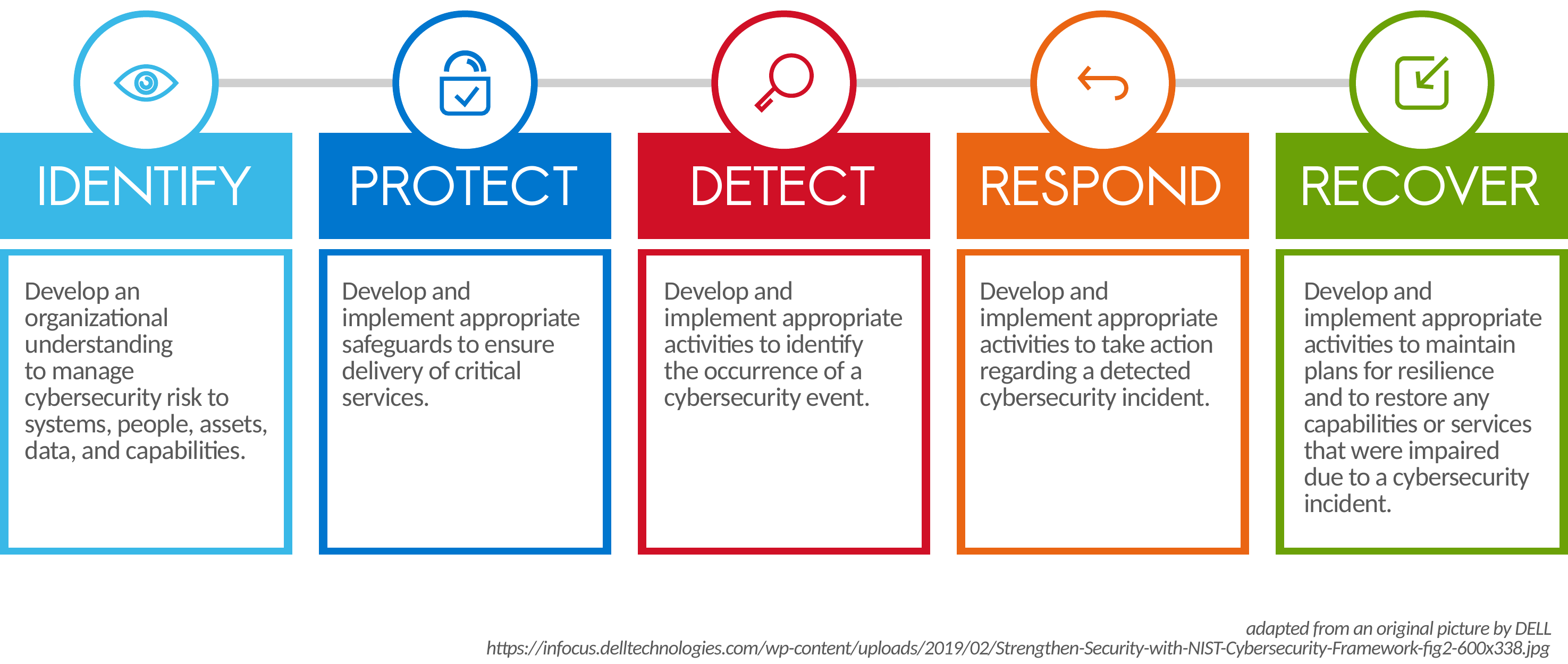}
\end{center}

In this section, we analyse how each function can be instantiated in the context of an organisation involved in the EV charging ecosystem.
Of course, this is a continuous process: new standards, protocols and policies are going to appear that regulate this ecosystem.
Nonetheless, the hardware shipped with chargers and cars that currently circulate is not obviously updatable to newer technologies or policies, and may become legacy.
For instance, the firmware and software of a legacy device cannot be updated to let the hardware work in the new required way.
For example, the devices might not have enough memory to store stronger secrets, or the processors may be not powerful enough to carry out stronger cryptographic operations, or else the communication channel cannot transfer more than some small amount of bits per seconds.

The (ENCS), in collaboration with ElaadNL, provides some guidelines for a future-proof design for EV charging devices~\cite{elaadnl2019cybersecurity}.
Security issues with legacy hardware and backward compatibility are not peculiar to the V2G ecosystem, but they represent a security dimension that we strongly recommend not to overlook.

\section{Identify}
\label{sec:nist-identify}

The {\bf Identify} function is perhaps the most crucial as it requires assessing the cybersecurity risk faced by the organisation.
The risk of an individual cybersecurity event is given by the likelihood of that event multiplied by its impact.
The risk faced by the organisation is the aggregation of the risk of all possible cybersecurity events.

The first challenge in computing risk is to list all possible events.
Unfortunately, there is no method that guarantees to be exhaustive, and the traditional approach consists in breaking down the problem in several categories of desired security properties such as the well-established {\bf CIA triad}: {\bf confidentiality}, {\bf integrity} and {\bf availability}.

\begin{wrapfigure}{r}{0.36\textwidth}
  \includegraphics[width=0.36\textwidth]{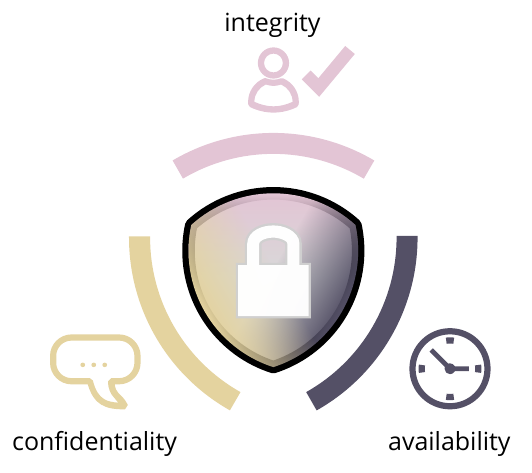}
\end{wrapfigure}
In a nutshell, {\em confidentiality} ensures that information shall only be readable by the intended recipients (which could include processing systems) and protects it from unauthorised third parties.
{\em Integrity} ensures that any modification can only be done by authorised agents.
{\em Availability} ensures the services offered by the system respond to the queries of the user within an expected time frame (i.e. available when needed).

Confidentiality, integrity and availability are core properties; yet there are additional desirable properties such as authentication, non-repudiation and auditability.
Authentication means that a particular identified party is not impersonating another.
Non-repudiation means that if some action is performed by an identified party (e.g. agreeing to the terms of a contract), this can be proven later and cannot be denied.
Auditability means that a trustworthy log describing how information has been used is available (e.g. important for forensic purposes)~\cite{fall2011tcp}.

Focusing on the core properties, the identification of cybersecurity events can therefore be done by first focusing on all events targeting confidentiality, then those targeting integrity, and finally those targeting availability.
More specific categories have been used in threat modelling, such as the STRIDE model.

It is necessary to first identify all potential attack vectors and attack surfaces. Some examples of attacks on EVs and EV chargers (and the issues fixed) are presented~\cite{tierney2020trust}.
Figure 3.1 illustrates attack vectors in the EV ecosystem.
Following this, a threat modelling~\cite{sae2016cybersecurity} exercise needs to be undertaken for every attack to identify  the various potential scenarios and threat actors, followed by an analysis of the impact of each threat.
The key challenges facing this complex ecosystem of EVs, V2G chargers and third party operators include:
\begin{itemize}
  \item physical limitations of devices and communication channels;
  \item heterogeneity, scale, and ad-hoc nature of threats;
  \item authentication and identity management;
  \item authorisation and access control; and
  \item implementation, updating, responsibility, and accountability~\cite{maple2017security}.
\end{itemize}

\begin{figure}[htbp]
  \caption{Possible attack vectors in the EV Ecosystem. Adapted from: Mike Nelson, DigiCert \& Oscar Marcia, Eonti Inc. - Public Key Infrastructure (PKI) for electric vehicles, \url{https://youtu.be/nEBJzPVZNd0}.}
  \label{fig:attack-vectors}
  \includegraphics[width=\textwidth]{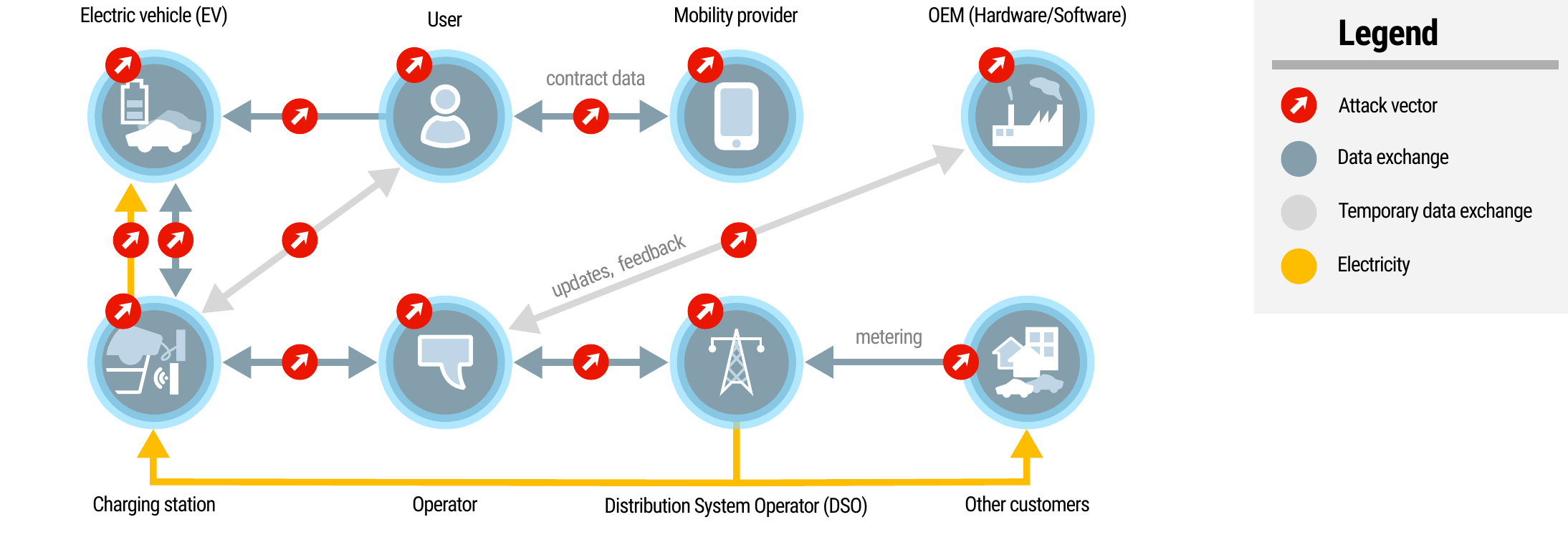}
\end{figure}
\subsection{Impact of cyber incidents}
\begin{wrapfigure}{r}{0.36\textwidth}
  \vspace{-2.4cm}
  \includegraphics[width=0.36\textwidth]{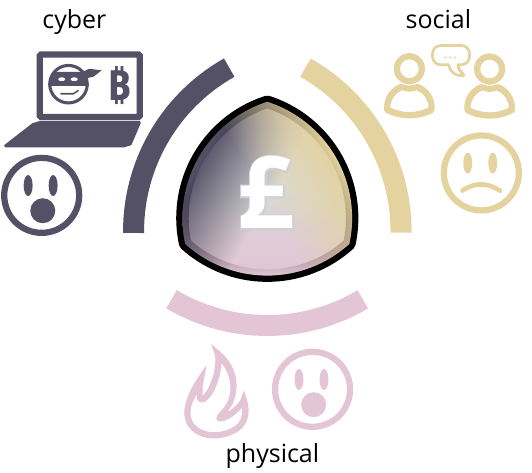}
\end{wrapfigure}
The second challenge in computing risk is to identify the likelihood and impact of a possible cybersecurity event.
Each event might expose the individual users, the company, their customer and the system.
The impact of such events could be generalised in three major areas: physical, social and cyber.
In general, the damages identified in those areas are eventually assessed in risks and severity reports, and relate to an economical or personal loss of the victim and to a benefit gained by the attacker.
So it is important to take into consideration the balance between the cost to carry out the attack and the benefit gained by its perpetrators, to avoid unnecessary costs of overly securing systems.

\subsubsection{Physical impact}
Attacks~\cite{atamli2020cyber,bellekens2020cybersecurity} that have a direct or indirect physical impact to victims can aim to
\begin{itemize}
  \item compromise devices through physical access,
  \item break hardware of devices, or
  \item provoke hazards that put in danger the health (or the life) of users.
\end{itemize}

\subsubsection{Cyber impact}
Attacks often aim to collect private data, e.g. encryption private keys, bank account numbers, contact lists, profile credentials and preferences, to be sold to companies that profit from such data or organise attacks on a larger scale.
In general, the cyber impact of such attacks does not break hardware but may leave the operating system in an unstable or ill state that requires manual intervention.

\subsubsection{Social impact}
Attacks can aim to slander or expose the private life of victims, or in the case of companies, to decrease the amount of trust among their actual or potential customers.
These kinds of attacks may simply manipulate the outcome of human behaviour (social hacking) without the need to directly exploit vulnerabilities of protocols and devices.

In the table below, we summarise possible attacks according to the security concepts of confidentiality, integrity and availability, and the impact of attacks in terms of physical impact, cyber impact or social impact.
An example from the table considers DoS attacks: these are attacks carried out with different techniques with the common goal to provoke a discontinuity or a delay in the service, thus undermining availability.
For the most part, the impact of such attacks to the communication network can put some devices in a state that requires a fresh restart or drain their battery, and can have an impact on the amount of trust on a company from the customers who can appreciate the reduction or lack of service.

\begin{center}
\smaller
\renewcommand{\arraystretch}{1.5}
\setlength\arrayrulewidth{2pt}
\arrayrulecolor{white}
\rowcolors{2}{oddrows}{evenrows}
\begin{longtable}{C|D|M|M|M|M|M|M|E}
  \caption{Possible security attacks classified according with the CIA triad and Impact.}
  \label{tbl:attack-type} \\
  & 
  & \thr[90]{Confidentiality}
  & \thr[90]{Integrity}
  & \thr[90]{Availability}
  & \thr[90]{Social}
  & \thr[90]{Cyber}
  & \thr[90]{Physical}
  &  \\
    \thh{Attack type}
  & \thh{Specific target (optional)}
  & \multicolumn{3}{c|}{\thh{Security}}
  & \multicolumn{3}{c|}{\thh{Impact}}
  & \thh{Example} \\
  \hline
  \endfirsthead

  \continuedtable{9} \\
    \thh{Possible Target}
  & \thh{Details}
  & \thh{C}
  & \thh{I}
  & \thh{A}
  & \thh{S}
  & \thh{C}
  & \thh{P}
  & \thh{Example} \\ \hline \endhead
  
  \continuetable{9} \\ \hline \endfoot
  \endlastfoot
  
  \thc{Denial of Service (DoS)} & 
    &       &       & $\rb$ & $\rb$ & $\rb$ &       & Communication broken either at aggregator level or broader \\ \hline
  \thc{Delay attack} & 
    &       &       & $\rb$ &       & $\rb$ & $\rb$ & Power requests at incorrect timing may cause breakdowns \\ \hline
  \thc{Replay attack} & 
    &       & $\rb$ &       &       & $\rb$ & $\rb$ & Inconveniently replaying power requests may cause breakdowns \\ \hline
  \thc{Snooping} & 
    & $\rb$ &       &       & $\rb$ &       & $\rb$ & Eavesdropping insecure communications may let the attacker link messages to the same user, therefore tracking, stalking, or stealing of private information \\ \hline
  \thc{Sybil attack} & Aggregator
    &       & $\rb$ &       & $\rb$ &       &       & Copy ID tokens to multiply energy charge for free \\ \hline
  \thc{Impersonation} & EV, aggregator
    &       & $\rb$ &       &       & $\rb$ &       & Steal energy from either directions \\ \hline
  \thc{Man-in-the-middle} & 
    & $\rb$ & $\rb$ &       &       & $\rb$ & $\rb$ & Messages can be tampered, i.e. charging decisions can be subverted \\ \hline
  \thc{Repudiation} & EV
    &       &       & $\rb$ &       & $\rb$ &       & Denial of power to legitimate EVs \\ \hline
  \thc{EV misbehave} & Aggregator, billing party
    &       & $\rb$ &       &       & $\rb$ & $\rb$ & Wrong information may entail wrong decisions \\ \hline
\end{longtable}
\end{center}

\section{Protect}
\label{sec:nist-protect}

The Protect function focuses on protecting product assets and information, hardware and software from cyber threats, including the cyber protection of  embedded electronic controller units (ECUs) in automotive electronics, securing  design of the architecture (segmentation, boundaries protection), and securing information like payment information and vehicle identification.

\subsubsection{Cyber}
Cyber security technical requirements should cover the supply chain.
For example, establishment of guidelines for cyber security during the procurement phase, to guide the customer through the security measures of the purchased EV or charge point.
The cyber security requirements for third parties involved in the smart or V2G charging process such as building an energy management system charging station should also be defined.
Different entities in the V2G ecosystem can store private information that belong to owners, e.g. the EV driver,  or customers, e.g. an aggregator or a charge point operator.
Companies may  store such information in databases and part of the information may be exchanged with other entities.

From the security perspective, the databases or files with private information need to be: {\bf stored in encrypted drives} to avoid information theft (think also of picking up a disposed charger), and {\bf be isolated/inaccessible} to other co-existing software.
The operating systems and firmware of devices shall enact standard security measures, as strong and random passwords (and not reconstructable from serial numbers) and avoid backdoors, adopt a proper implementation of access control of files and other resources, and they must allow updates from verified sources, e.g. signed by manufacturers.
When part of such information needs to be exchanged, the entities shall first {\bf establish a secure channel}, e.g. through secure protocols as TLS, then exchange messages securely.
Security of communication is discussed more in detail in the next dedicated section.

Procurement of third-party components can also be logged into a Digital Bill of Materials (DBOM).
The diversity of the global automotive and energy supply chain makes it difficult to know the provenance of every chipset, device and connected component required.
Furthermore, the current state of the supply chain does not provide information on the buyer, the assembler and the reseller.
In an ever more connected world, such information is critical to prevent cyber-attacks and identify threats before an asset is on the market.
DBOMs can take the form of a shared ledger with or without a central authority collecting information on the journey of every component ranging from chipsets  and cargo identifiers to the company assembling the smart chargers.
It is therefore key for the DBOM to be immutable while providing adequate information for traceability.
The same concept can be applied for the software written for every component ranging from firmware to the user interface through a Software Bill of Materials (SBOM).

\subsubsection{Physical}
Physical security is also considered to be a concern in the V2G charging process.
Several devices are physically exposed to potential attackers, e.g. cars are parked in public spaces and some of the chargers and charging cables are publicly placed (e.g. parking lots, kerbside parking) with little or no extensive physical security measures applied .

The internal memories and devices must be protected against cloning and tampering at least in the same way we protect electrical boxes on public spaces.
Managed EV charging technologies such as V2G adds an additional security requirement, as private information will be exchanged between vehicles, chargers, and other entities.
Thus, the communication links need to be protected as well as the power links.
Particular attention is required if and when wireless charging is introduced to ensure that cars are protected from being discharged by bogus or cloned chargers, as they would not need to access a cabled outlet (that could be key-locked).

Cyber security requirements must consider all entities that are involved in this process, starting from the EV charge point but also the customer, the EV, the charge point operator and other third party operators such as electric companies involved in the integration of electric vehicles into the grid.

The Protect function also includes part of the design of both power and communication networks, and strongly refers to the risk and security requirements covered by the Identify function.
It relates to the system configuration in terms of their capability and their topological or geographical displacement.
At this level, it is important to implement what is required to provide availability, e.g. redundant power and communication links or distributed databases, as well as what physical security measures and what communication protocols are best suited to protect confidentiality and integrity of exchanged messages or stored secrets.
Cybersecurity analysis at this level can be done with the aid of tools.
\textit{\textbf{Literature counts several of such tools that can be applied to analyse the EV charging infrastructure:}} they are usually extensions or bespoke applications of tools previously used to analyse cybersecurity of industrial systems connected to both the Internet and the power grid.
They can be used to analyse manually designed attack scenarios and, to some extent, incident responses.
Those tools can help only when designing the architecture, but can hardly capture vulnerabilities of a deployed system as it cannot capture all its aspects.
It is best practice to establish at this level what resources are required for a future-proof device in terms of resources to allow for security updates and other compliance updates that can relate to foreseeable evolution of policies.
An example of a certification programme is the ENERGY STAR\textregistered Program Requirements for Electric Vehicle Supply Equipment~\cite{energy2017energy}.

Once the architecture is decided, next is its implementation, i.e. the practical construction of devices (hardware) and the implementation of algorithms and protocols (software) both respecting regulations imposed by governmental, international or local policies and standards.
Security practices at this stage are the analysis of the composition of security protocols, their implementation, and the security of both device systems\footnote{It would be pointless to use any form of encryption in communications between (flawed) systems that allow attackers to retrieve private keys.} and the communication networks.
As formal verification of implementations is very complex and time consuming, security best practices here are often empirical: penetration testing, constant updating of operating systems (and firmwares) and their configuration against the latest security recommendations, e.g. by NIST.

\section{Detect, Respond, and Recover}
\label{sec:nist-detect-respond-recover}

The EU Security of Network \& Information Systems Regulations (NIS Regulations) have been introduced in the UK in 2018.
It has been reviewed in May 2020, where they could preliminary assess that it incentivised operators to improve their security but long-term benefits are still to be understood.
It provides legal measures to boost the overall level of security (both cyber and physical resilience) of network and information systems that are critical for the provision of the essential services of transport and energy among others.

In particular, they establish enforcement and penalties for operators of essential services not matching appropriate security measures.
Directly citing from them, the operator must take appropriate and proportionate measures to prevent and minimise the impact of incidents affecting the security of the network and information systems used for the provision of an essential service, with a view to ensuring the continuity of those services~\cite{uk2018network}.

The three functions of Detect, Respond and Recover are strictly related and include all the necessary tasks required to keep security and incident-responses active and effective once the whole architecture is up and running.

Detecting cyber-threats against electrical vehicles and the infrastructure is an herculean task given the large spread of the attack surface~\cite{wolf2004security}.
Integrating cyber-security to an existing and evolving infrastructure requires a considerable amount of operational planning.

From a human perspective cyber-ranges can be used to simulate a representative infrastructure and associated cyber-attacks against the power grid to evaluate and train operators to identify, detect and respond in a timely manner.
From an end-user perspective it is essential to educate the users to a variety of risks, which remain to be identified (e.g. Connecting a vehicle to open wireless hotspots in the vicinity of charging stations).

From a technical standpoint, EV and the grid must encompass a range of traditional  measures and protection devices such as firewalls, intrusion detection systems, intrusion prevention systems, disabling default credentials, encryption of communications, etc.
Emerging techniques such as digital twinning to monitor the state of the system or cyber-deception can also be used to trick the attacker into believing they are interacting with a real system while providing them with false information.
Cyber-deception also allows for a low false positive rate and modus operandi information gathering for the recovery phase\footnote{\url{https://www.lupovis.io}}.
Information from the various cyber-security devices should be aggregated through a Security information and event management (SIEM) for analysis by a security operator.

Alongside the deployment of cyber-security components, the  Techniques, Tactics and Procedures (TTPs) of attackers should also be analysed and conclusions drawn from previous attacks.
This can be achieved by MITRE's Adversarial Tactics, Techniques and Common Knowledge (ATT\&CK®) framework.
\begin{figure}
  \caption{A screenshot of the ATT\&CK Matrix for Enterprise.}
  \label{fig:mitre-matrix}
  \begin{center}
    \includegraphics[width=\textwidth]{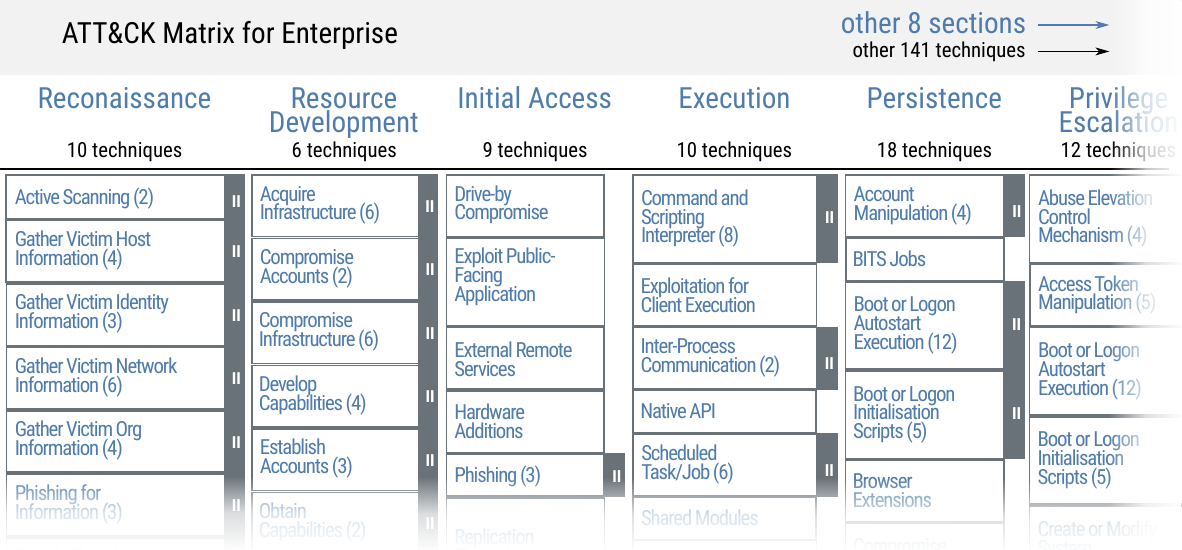}
  \end{center}
\end{figure}
As shown in Figure~\ref{fig:mitre-matrix}, a list of TTPs employed by attackers in actual attacks over Enterprise systems (note there is also a {\em mobile} structure in the framework).
The ATT\&CK matrix provides information on the modus operandi of attackers to achieve persistence and disrupt an infrastructure.
The matrix further allows to identify critical components and provide guidelines on ``why'', ``how'', and ``what'' to protect to alleviate a cyber attack.
It is worth mentioning that ATT\&CK lists only reported incidents, so security officers and administrators should take other protective measures to complement their approaches.

Establishing testing phases through penetration testing is also good practice to evaluate the security of services and components however, it is key to understand the potential consequences of the test.
For example, it is recommended to test critical operational components offline as testing may lead to undefined behaviour under stress.

Quick recovery from a cyber-attack is essential and can be planned with the cyber recovery operational framework.
\begin{figure}
  \caption{Cyber recovery operational framework. Links emphasise the most relating or interdepending points.}
  \label{fig:cyber-recovery-operational-framework}
  \begin{center}
    \includegraphics[width=\textwidth]{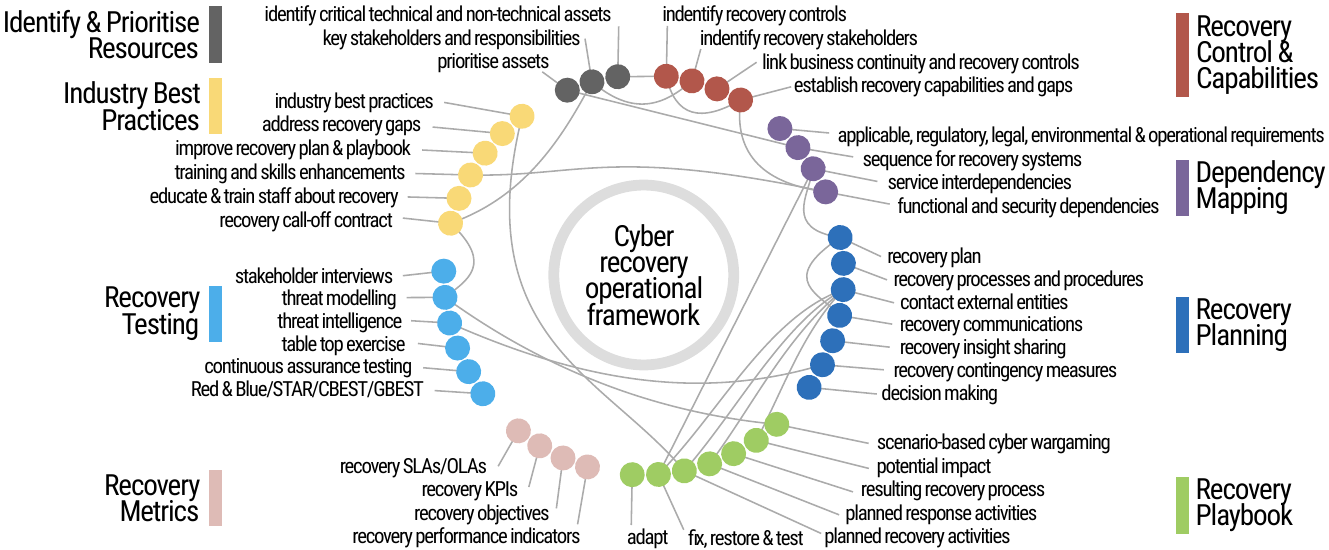}
  \end{center}
\end{figure}
Figure~\ref{fig:cyber-recovery-operational-framework} depicts the cyber recovery operational framework which indicates the step to take ahead and during a cyber attack to recover the infrastructure quickly while ensuring business continuity~\cite{cmric2020cyber}.

After a cyber-attack, it is essential to collect information on the attacker to quickly respond to the following question  ``who'', ``what'', ``when'' and ``where'' through digital forensics.

During the digital forensics procedure it is essential to follow the ACPO guidelines~\cite{athena2020acpo} which provide a set of guidelines for collecting digital evidence after a crime.
Information can be gathered from the intrusion detection systems, firewalls and cyber-deception tooling as well as the Security Information and Event Management (SIEM).
Furthermore, Information from the physical components such as the charging stations and other components involved in the cyber-attack should also be secured.

As a final remark, the V2G technology is expected to coexist along with other technologies to form smart grids and smart cities which are in constant evolution.
As such, revisiting each and all phases is required to keep the infrastructure in a secure status.

 \chapter[Privacy in the EV Charging Infrastructure]{\nohyphens{Privacy in the EV Charging \tnl Infrastructure}}
\label{ch:privacy-ev-charging-infrastructure}

In addition to ensuring the security of the EV ecosystem, it is of utmost importance to consider data privacy.
In this section, we discuss data privacy in the context of IoT devices that could be applied to V2G.

V2G services have the potential to add up to the already existing {\em unrelated} services offered to end-users through the Internet, e.g. directions on maps.

Users of Internet services, such as directions on maps, familiarised themselves with the tradeoff of providing private data (i.e. geolocation) in return of an improved personalised service (e.g. traffic-aware directions, locations of nearby restaurants, etc.).
Any restriction enforced by the user, e.g. disabling location sharing will be reflected in either the impossibility to use some service or the limited quality of the same.

In a similar way, there is a tradeoff involved if EV users participate in V2G services to reduce their transport cost.
To allow an optimal V2G services, users would trust the V2G operator to manage their charging events and track their driving patterns (i.e. Time of arrival, departure).
There is also personally identifiable information (PII) involved such as the name, banking info.
Consequently protecting V2G communication is a key requirement, including security of databases or cache memories storing private information.

\begin{wrapfigure}{r}{0.5\textwidth}
  \includegraphics[width=0.5\textwidth]{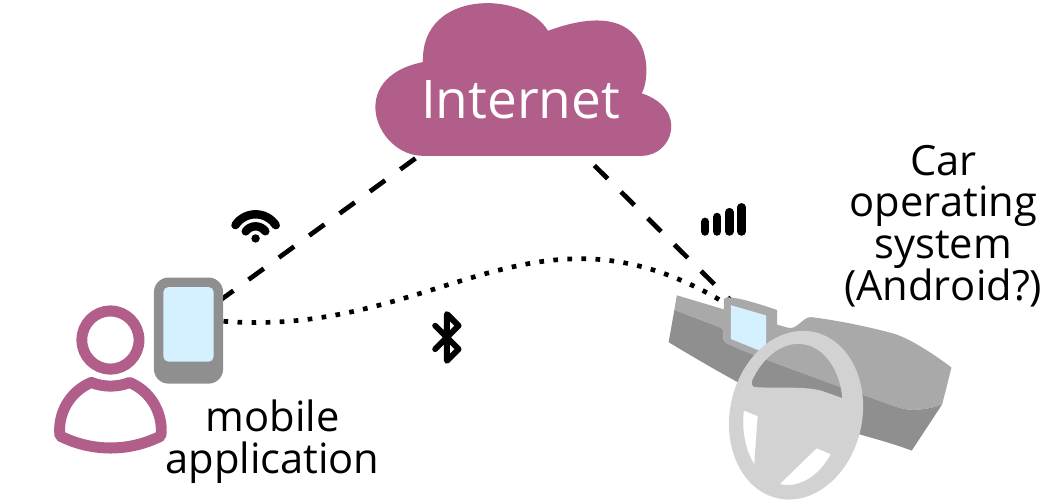}
\end{wrapfigure}
Privacy becomes even more important when the operating system of a car would be linked to the operating system of a phone and allows for augmented monitoring and control of the users' activity.
For example, the car/charger can extract the calendar information about the time and destination of upcoming trips from the user's phone and accordingly set the charging profile.

Depending on the vehicle and manufacturer, the personal data required to make the services work varies.
For example, most mobile applications by manufacturers (e.g. Tesla, Audi, Toyota and others) require read-and-write permissions to the user's {\em calendar}.
An important caveat on {\bf understanding permissions} in mobile applications is that, even though some permissions are required to perform operations, that does not imply that all permitted operations will be actually done: for example, if an application is granted calendar access, it will {\em unlikely} release all your schedule to some third party by email without your knowledge, yet it means it might.
This reconciles with the importance of using open source applications, where anybody can potentially inspect the source code and see what for or where private information is collected and under what circumstances.
We strongly recommend the anonymisation and aggregation of data to protect users' privacy.

Sharing users' data by default without engaging the user is risky because they might resist participating in smart charging and V2G initiatives.
Some users, indeed, distrust ``smart'' systems and could resist migrating to them.
To draw a comparison to smart meters roll-out in several countries, previous work indicated that overlooked non-technical factors such as social and ethical considerations could have a considerable impact on the success of the smart meter roll-out and a future smart energy system.

\section{Existing Privacy Regulations and Guidance}
\label{sec:privacy-regulations}

In addition to gaining users' trust, new government regulation and guidance emphasise the importance of data privacy requirements, which should be applied to connected smart charging systems.
Regulation and guidance include the GDPR regulation; the guidance on Internet of Things (IoT) and the guidance on Principles of cyber security for connected and automated vehicles.

First, GDPR (General Data Protection Regulation) was enforced May 2018, with stronger rights for end-users with respect to their personal data\cite{ico2018guide}.
Specifically, the users have the right to: be {\bf informed}, {\bf access}, {\bf rectify} or {\bf erase} their data, {\bf restrict} or {\bf object} processing, data portability and other rights related to {\bf automated decision making} and {\bf profiling}.

These end-users' rights detailed in GDPR are emphasised in a government publication on the Internet of Things (IoT)~\cite{department2019secure}.
The document highlights the requirement to ensure that personal data is processed and protected in accordance with data protection law.

The document indicates that device manufacturers and IoT service providers must provide consumers with clear and transparent information about how their data is being used, by whom, and for what purposes, for each device and service.
Moreover, users should be provided by means to preserve their privacy by configuring device and service functionality appropriately.

In more detail, where personal data is processed on the basis of consumers' consent, this must be validly and lawfully obtained, with those consumers being given the opportunity to withdraw it at any time.
Consumers should also be provided with guidance on how to securely set up their device, as well as how they may eventually securely dispose of it.

While the document does not specify electric vehicles, EV charging would fit the definition of a connected IoT device and the actors involved in EV charging should adhere to this government's code of practice for consumer internet of things.
In particular and when applicable, EV ecosystem actors would need to be mindful of:
\begin{shadequote}{}
Providing clear and transparent information to consumers about what personal data devices and services process, the organisations that process this data, and the lawful basis on which the processing takes place.
\end{shadequote}

\begin{shadequote}{}
Building privacy and security into the product lifecycle from the design phase, and ensure these are continued throughout.
\end{shadequote}

\begin{shadequote}{}
Ensuring that appropriate technical and organisational measures are in place to protect any personal data, including processes to ensure the confidentiality, integrity, availability and resilience of processing systems and services, and regular testing to ensure the effectiveness of such measures.
\end{shadequote}

The Information Commissioner's Office (ICO) is the UK's data protection regulator, providing advice and guidance to organisations and consumers and, where necessary, undertaking appropriate and proportionate enforcement action.
For example, ICO fined Talk Talk a penalty of £400,000 for failing to properly protect customer data from a cyber-attack.
ICO's investigation found that the attack on the company could have been prevented in Talk Talk has taken basic steps to protect customers' information\footnote{\url{https://ico.org.uk/about-the-ico/news-and-events/news-and-blogs/2016/10/talktalk-gets-record-400-000-fine-for-failing-to-prevent-october-2015-attack/} [last retrieved 2019-01-11].}.
More specifically to the automotive sector, DfT in partnership with the Centre for the Protection of National Infrastructure (CPNI) and the Centre for Connected and Autonomous Vehicles published 8 principles for obtaining good cyber security within the automotive sector including.
Principle 7 relates to data privacy and states that: ``the storage and transmission of data is secure and can be controlled''.
In detail:
\begin{shadequote}{Principle 7.1}
Data must be sufficiently secure (confidentiality and integrity) when stored and transmitted so that only the intended recipient or system functions are able to receive and/or access it.
Incoming communications are treated as unsecure until validated.
\end{shadequote}

\begin{shadequote}{Principle 7.2}
Personally identifiable data must be managed appropriately.
This includes:
\begin{itemize}
  \item what is stored (both on and off the ITS / CAV system)
  \item what is transmitted
  \item how it is used
  \item the control the data owner has over these processes
  \item where possible, data that is sent to other systems is sanitised.
\end{itemize}
\end{shadequote}

\begin{shadequote}{Principle 7.3}
  Users are able to delete sensitive data held on systems and connected systems.
\end{shadequote}

Principle 7 of the guidance on the key principles of vehicle cyber security for connected and automated vehicles is aligned with the text of GDPR and the guidance on IoT.
This would suggest that relevant actors in the EV ecosystem would need to be mindful of data privacy issues, with the ICO, data regulator, providing guidance and when necessary enforcing fines for failure to comply with data protection laws.

As a final remark on the near future of regulation, the publication of the European Cyber Security Act has triggered the work on the Network Code on energy-specific cyber security which is currently being developed.
Network Codes have legal status within the European Union, and the UK has implemented
previous codes.
 \chapter[Communication protocols linking various EV ecosystem entities]{\nohyphens{Communication protocols linking \tnl various EV ecosystem entities}}
\label{ch:protocols}

Communication protocols provide a set of rules and guidelines to facilitate communication and data exchange between two or more entities.
A protocol would define the interface between two or more interacting entities to ensure compatibility between these different systems~\cite{ferwerda2018advancing}.

\begin{figure}[htbp]
  \caption{Various EV ecosystem entities connected by communication protocols - more information on entities and protocols can be found at \url{https://energyinformatics.springeropen.com/articles/10.1186/s42162-020-0103-1}.}
  \label{fig:v2g-ecosystem-information-flow-horizontal}
  \begin{center}
    \includegraphics[width=\textwidth]{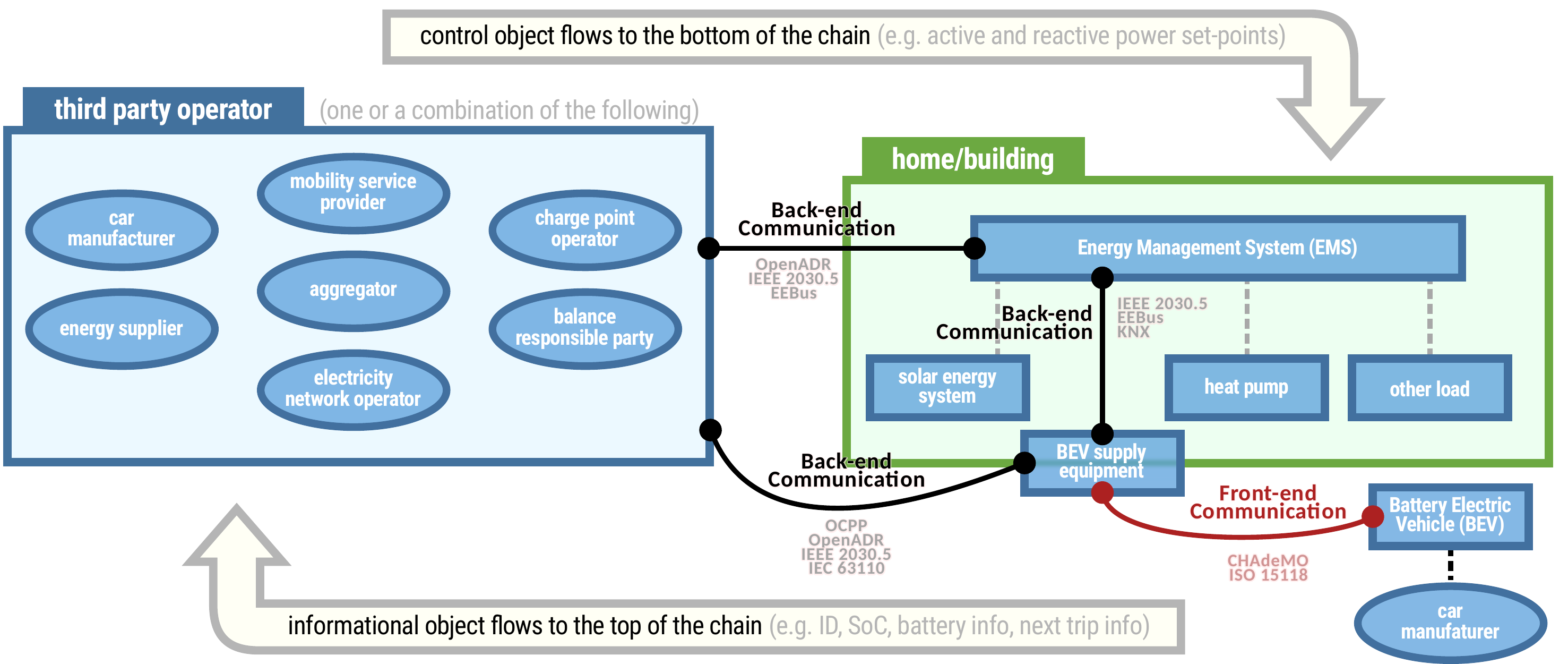}
  \end{center}
\end{figure}

Management of BEV charging requires increasing need for coordination and communication between various mobility and energy entities, see Figure~\ref{fig:v2g-ecosystem-information-flow-horizontal}.
Specifically, informational and control objects need to be exchanged between BEV chargers and entities to allow electricity system support, such as electricity demand peak reduction.
Information exchanged such as car identification, battery state of charge (SoC), battery size, energy required for the next trip would flow up across entities.
Based on that information and on information collected from the electricity system infrastructure (e.g. frequency, current, and voltage data), chargers' power set points are determined and sent down to control the charging process, ensuring proper integration of vehicles into the electricity grid.

Communication protocols linking various entities in the EV ecosystem can be divided into front-end and back-end protocols~\cite{schmutzler2013evaluation}.

Front-end protocols define the link between car and charge point and specify requirements for plugs, charging topologies (on-board/off-board charging equipment; conductive/inductive charging), communication, safety and cyber-security.
Protocols such as CHAdeMO and ISO15118-20 allow bidirectional power flow (i.e. V2G) between the car and a charger.

Back-end protocols, emphasising communication and cyber-security requirements, define the link between charge point and a third party operator.
Some of these open protocols are OCPP, IEC63110, OpenADR, EEBus and IEEE2030.5.

A brief security review of some of these protocols is presented below.
We start with two front-end protocols CHAdeMO and ISO15118, then we review four back-end protocols, namely OCPP, OpenADR, IEEE 2030.5 and EEBus.

\section{Front-end Protocols: EV to Charger}
\label{sec:front-end-protocols}

\subsection{CHAdeMO}
CHAdeMO is a DC charging protocol enabling V2G operations for electric vehicles, i.e. charging and discharging of their internal battery.
CHAdeMO lacks secure communication features and it relies on CAN communication which is described below.

Vehicles that plug to chargers with CHAdeMO expose their CAN bus to the (untrusted) charger.
Therefore, security concerns on the CHAdeMO are mostly related to the unencrypted communication through the CAN bus, and in particular to the ability of a malicious entity to control or program other ECUs exposed by the CAN bus.
CHAdeMO follows the standard IEC 61851-23:2014 for the actual charging and discharging operations, and follows the standard IEC 61851-24:2014 for the digital communication between the car and the charger.

In 2019, CHAdeMO announced that it is co-developing the next generation ultra-fast EV charging standard called ChaoJi in collaboration with China Electricity Council (CEC) who is behind the GB/T standard for EV charging in China.

\subsubsection{Security of ChaoJi/CHAdeMO 3.0}
Currently, ChaoJi is not implementing any security for the communication and will still communicate using the CAN bus (not only for retro-compatibility purposes), following the IEC 61851-24.
However, they recognise the importance of security and plan towards a unified communication protocol that would run over Ethernet and TCP/IP, and would follow a PKI infrastructure and employ TLS, similarly to how TLS is already used in the Internet.

\subsubsection{CAN bus}
CAN (Controller Area Network) is a message-based protocol introduced in the '80s by Robert BOSCH GmbH that allows electronic components of a car to communicate with each other's applications without requiring a central computer.
Being a low-level protocol, messages are exchanged by ECUs unencrypted; CAN does not support any security features.
In most implementations, applications are expected to deploy their own security mechanisms.

Some functions that use the CAN bus are safety-critical, e.g. firmware updates or brakes, and are protected with passwords.
However, modern cars are equipped with hundreds of ECUs, and their coexistence with TCUs (Telematic Control Units), connected to the Internet, increased the necessity of strong security guarantees.
\begin{center}
  \includegraphics[width=\textwidth]{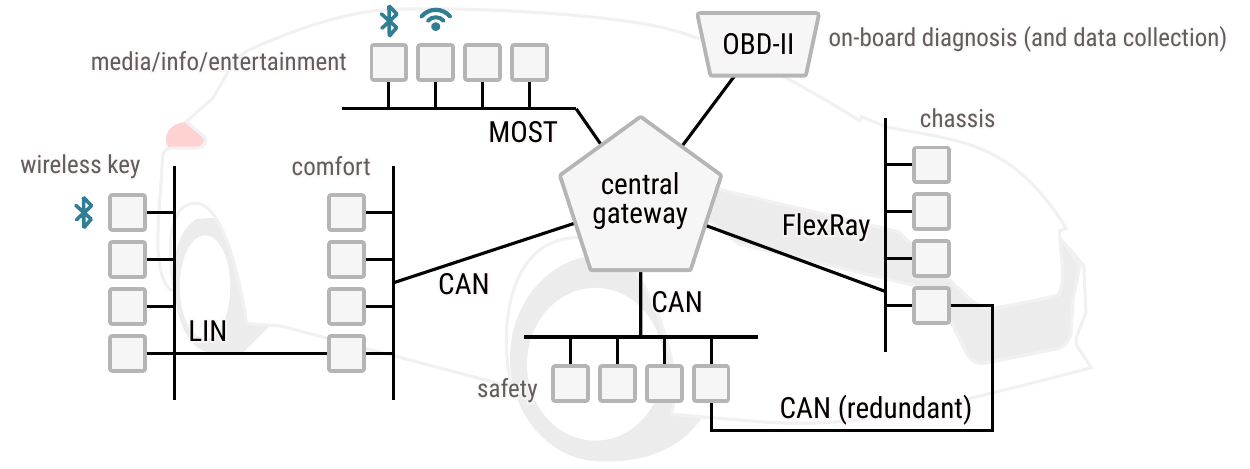}
\end{center}
Numerous examples of security concerns and attacks on the CAN have been described~\cite{koscher2010experimental,cho2017viden,pese2019librecan,urquhart2019cyber,frassinelli2020know,islam2020improving}.
A devastating remote attack on Jeep shown by Miller and Valasek~\cite{miller2014survey} got the attention of companies on the seriousness of CAN bus attacks, even though details of previous attacks on different vehicles were already published~\cite{pagliery2013tesla,cbs2015hacked}.

\subsection{ISO 15118}
ISO 15118 is another front end protocol connecting electric vehicles to charging infrastructure with some additional capabilities such as Plug\&Charge and extensive security measures.
One main feature of ISO 15118 is using digital certificates\footnote{More information on certificates in Section~\ref{sec:pki}.} to secure the communication.
The protocol also allows automated authentication and authorization.
ISO 15118 is part of the Combined Charging System (CCS), which is a set of hardware and software standards for charging systems.

ISO 15118-20 is currently under development, expected by the end of 2021, and will define a use case for bidirectional power transfer.
There already exist demonstrations using ISO 15118-20 for V2G (e.g. Renault).

\subsubsection{Security}
An important difference between CHAdeMO and ISO 15118 is that the former makes direct use of the CAN bus and ECUs internal to the car, while the latter establishes the requirements of the Physical and the Data link layers.
Also ISO15118 {\em suggests}\footnote{"TLS is not mandatory for certain Identification Modes other than the Plug-and-Charge Identification Mode", from ISO15118-2:2014 (revised and reconfirmed in 2020).} the adoption mature security protocols, i.e. TLS\footnote{TLS is a well studied suite of security protocols widely used in the Internet, whose most popular network model is TCP/IP. TLS is implemented at the application layer of TCP/IP.} while CHAdeMO doesn't (but expected in future versions of its successor ChaoJi).

\begin{figure}[htbp]
  \caption{Reference to the OSI model of CHAdeMO, ISO 15118, and the popular TCP/IP.}
  \label{fig:osi-tcp-can-chademo-iso15118}
  \begin{center}
    \includegraphics[width=.8\textwidth]{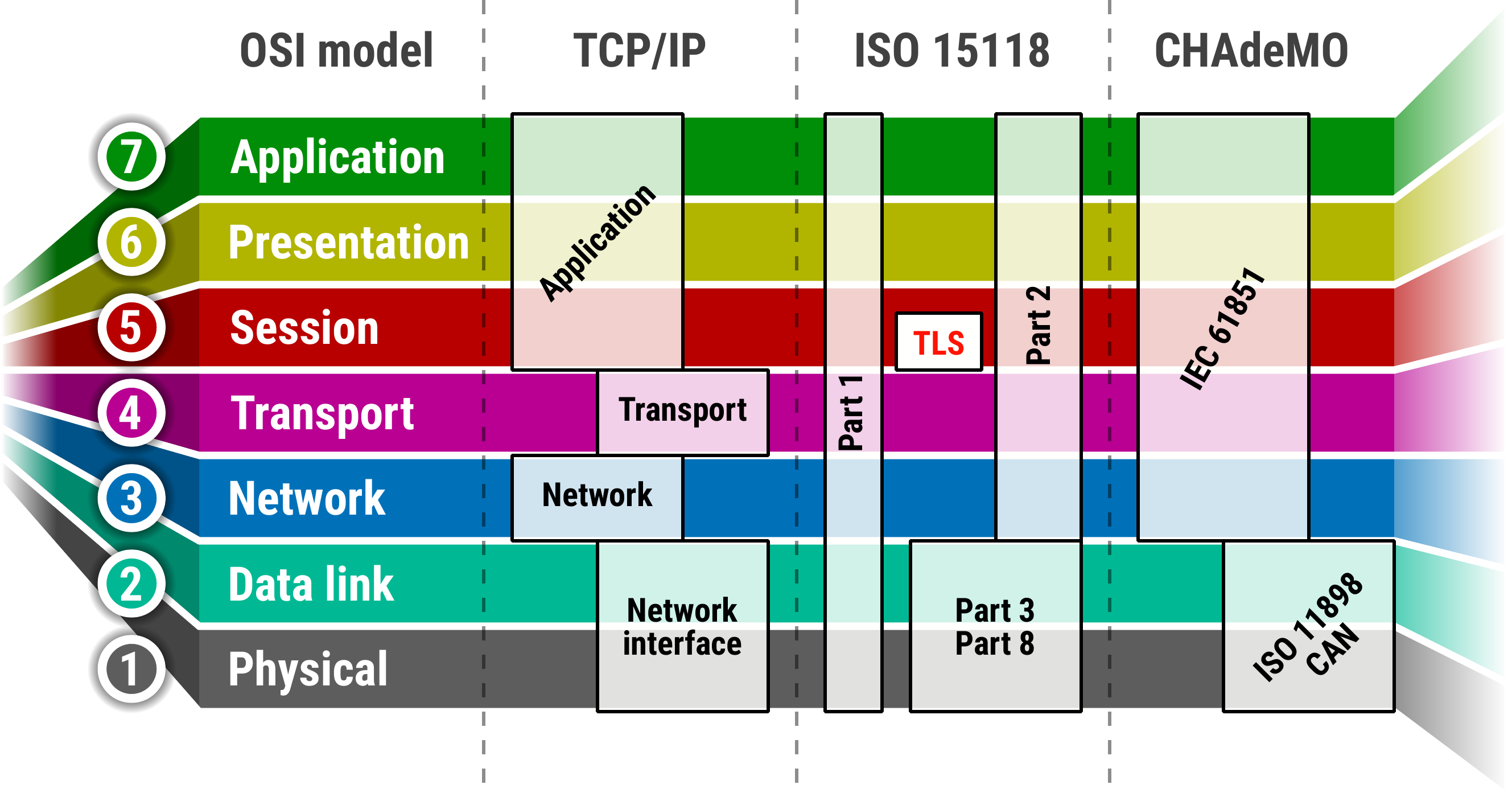}
  \end{center}
\end{figure}

ISO 15118~\cite{multin2019meet} is organised in parts.
Comparable parts with CHAdeMO are summarised in Table~\ref{tbl:chademo-vs-iso15118} and overlaid on top of the Open System Interconnection model (OSI) as specified in the ISO 15118 documentations, Figure~\ref{fig:osi-tcp-can-chademo-iso15118}.
OSI is the traditional model of reference for communication~\cite{zimmermann1980osi}; the table also shows the TCP/IP protocol suite as it is the most commonly used in the Internet.

It is worth noting that the ISO 15118 documentation places the TLS protocol at the transport layer, However, TLS is a protocol that comes on top of the transport layer~\cite{rescorla2008transport}.
\begin{center}
\small
\renewcommand{\arraystretch}{1.5}
\setlength\arrayrulewidth{2pt}
\arrayrulecolor{white}
\rowcolors{2}{oddrows}{oddrows}
\begin{longtable}{l|c|c|c|c}
  \caption{Comparison of standards for CHAdeMO (IEC 61851) and ISO 15118.}
  \label{tbl:chademo-vs-iso15118} \\
  & \multicolumn{2}{c|}{\thh{Energy transfer}} & \multicolumn{2}{c}{\thh{Communication}} \\
  & \thh{Wired}           & \thh{Wireless}        & \thh{Wired}         & \thh{Wireless}      \\
  \hline
  \endfirsthead

  \continuedtable{5} \\
  & \multicolumn{2}{c}{\thh{Energy transfer}} & \multicolumn{2}{c}{\thh{Communication}} \\
  & \thh{Wired}           & \thh{Wireless}        & \thh{Wired}         & \thh{Wireless}      \\ \hline \endhead
  
  \continuetable{5} \\ \hline \endfoot
  \endlastfoot
  
  \thh{CHAdeMO}   & IEC 61851-23            & --                      & IEC 61851-24          & --                    \\ \hline
  \thh{ISO 15118} & ISO 15118-3,1,2         & ISO 15118-8,1,2         & ISO 15118-3           & ISO 15118-8           \\
\end{longtable}
\end{center}

\subsection{Security of Proprietary protocols: the example of Tesla}
Tesla built their own alternative for high-level communication between electric vehicle and charging station.
The charger uses a protocol over the CAN bus, as found out by hacks in reverse engineering fashion~\cite{mahaffey2015hacking}.
Perhaps the most known vulnerability of Tesla is from Nie et al.~\cite{nie2017free}.
They describe how they managed to inject malicious messages into the CAN bus passing through the wireless communication, exploiting an anomaly in the gateway that is connected to both the Ethernet and the CAN bus.
Their attack was very serious as they could gain remote control of the car.
Interestingly, the vulnerability assessed by Nie et al. has been answered with an update from Tesla in only 10 days.

\section{Back-end Protocols: Charger to third party operators}
\label{sec:back-end-protocols}

\subsection{OCPP}
The Open Charge Point Protocol (OCPP)~\cite{leeuw2019meet}  defines communication protocols between chargers and central management systems and it is becoming the de facto standard.
Up to version 1.6, security was somehow overlooked, but the Open Charge Alliance published a security White Paper to complement the protocol applying the security of version 2.0 to 1.6-J~\cite{open2018enhanced}.
OCPP v2.0.x includes secure firmware updates, security logging and event notification, and security profiles for authentication and secure communication.
OCPP uses TLS to achieve security, and explicitly design key management for client-side certificates.
OCPP 2.0 supports ISO 15118, smart charging and PnC functionalities.
Using a combination of OCPP 2.0 and ISO 15118 with TLS would ensure a secure communication between EVs, chargers and central management systems.

\subsection{IEEE 2030.5}
The IEEE 2030.5~\cite{lum2020meet} standard is a suite of communication protocols to connect and directly control devices.
Its design is oriented and optimised for home area network devices.

Relating to the V2G ecosystem, it allows communication between any of the most relevant entities: aggregators, home-smart devices, chargers, EV (includes EV-charger communication).

We particularly note that IEEE 2030.5 put security as a priority in their design, especially if compared to their predecessors or standards they build on top of (e.g. IEC 61850).
To authenticate entities, IEEE 2030.5 describes life-long certificates and does not cope with EV mobility (after all, the home EV charger is not currently shared with strangers).
IEEE 2030.5 does not support a PKI: they actually go the opposite direction and provide life-long certificates that cannot be updated or revoked\footnote{Revocation is a very important feature! For example, due to a bug discovered on TLS in 2014, a considerably large number of affected websites needed to revoke and reissue their certificates.} and must be known privately.
Thus, despite its strong security guidelines \textbf{\textit{(e.g. follow NIST guidelines for encrypting communications using cipher suite.),}} it might not be easy to extend it to support a large scale deployment of an EV charging infrastructure (i.e. the protocol focus on communication of the devices in a small area (house or buildings) so a PKI infrastructure wasn't used, although it would be needed to extend to regional and national scale connected infrastructure.

\subsection{OpenADR}
Open Automated Demand Response (OpenADR)~\cite{bienert2020meet}  is an open information exchange model for distributed energy resources (DERs).
Their model typically relies on a gateway device, or aggregator to translate utility Demand Response (DR) and DER requirements into specific device behaviours.

OpenADR mandates the usage of TLS for mutual authentication, message integrity and confidentiality.
This may be an issue for devices that have been already manufactured with limited resources (e.g. insufficient computing power or communication bandwidth).
Moreover, their model specifically requires the usage of a PKI, where trusted certificate authorities (CAs)\footnote{More information on CAs in Section~\ref{sec:pki}.} sign certificates for each and all nodes, and authentication is achieved using those certificates.

OpenADR sees its nodes as virtual nodes (their internal architecture is abstracted) and this approach is much less direct toward device control than the approach of IEEE 2030.5.
Even if some of the functionality offered by OpenADR and IEEE 2030.5 overlap, one does not necessarily exclude the other on the same communication network.

\subsection{EEBus}
EEBus~\cite{fiege2020meet} is a suite of protocols that aim to harmonise communications in the Internet of Things (IoT) and has been adopted by many companies; in particular, they focus on the data structure and communication exchanged between entities.
They specify several protocols at different layers of communication: the SPINE protocol at the application layer (if compared to the OSI model) and the SHIP protocol at the network layer.
SPINE is security-agnostic, i.e. does not consider security if not for the trust that entities are supposed to have with each other to exchange data in the first place.
To achieve security, SPINE can be run over SHIP, which uses TLS and aims to provide a secure TCP/IP-based solution.
As such, EEBus is suitable to run over the Internet and can be, in principle, used for the secure communications of EV charging infrastructure. 

EEBus is open and free to the public \footnote{Freely accessible specifications: \url{https://www.eebus.org/}}, and it does not yet provide a royalty-free implementation (i.e. the specification is free but the actual implementation(i.e. the code) is not freely available. \footnote{\url{https://medium.com/grandcentrix/will-microsoft-joining-eebus-finally-bring-us-an-open-source-reference-implementation-3432a93dd2e6}}.

\section{Public Key Infrastructure}
\label{sec:pki}

The communication infrastructure required for the V2G ecosystem to be enacted is similar to the Internet, where we already securely exchange confidential messages, sign documents, and authenticate users and companies with the aid of a Public Key Infrastructure (PKI).
We discussed several communication protocols required to link various entities in the EV ecosystem and a PKI is a key component for the secure implementation of some of these protocols (e.g. ISO15118-20; OCPP 2.0; OpenADR).
The following paragraphs briefly introduce the general concepts behind PKI, then a PKI in the context of the EV ecosystem.

First, we briefly describe the basic cryptographic structures~\cite{fall2011tcp,kobeissi2020modern}
\footnote{See also Khan Academy-Cryptography\url{https://www.khanacademy.org/computing/computers-and-internet}} at the basis of PKIs in order to secure systems (i.e. to ensure confidentiality and authentication).
The mathematical instruments provided by cryptography include enciphering algorithms, key agreement algorithms and signature algorithms.
They provide a sound technical basis for constructing secure communications; on this basis, their widespread deployment across multiple administrative real-world domains requires more work (e.g. PKI).

Encryption algorithms can be based either on {\bf symmetric} keys or on {\bf asymmetric} keys.
A {\em key} is a particular sequence of bits required to perform {\bf encryption} or {\bf decryption} operations.
Without the right decryption key, it becomes extremely difficult to retrieve the plaintext of a given ciphertext.

Symmetric-key cryptography is also called {\bf private-key} cryptography as {\em one} private key that can be used for both encryption and decryption is generated.
Exchanging the key between entities requires a private channel.

Asymmetric-key cryptography is also called {\bf public-key} cryptography, as public keys are generated along private keys.
The public key can be used in encryption algorithms by anybody, and there is no need to exchange keys through private channels.
As a drawback, public-key algorithms are computationally more expensive than private-key algorithms; consequently, most systems today use {\bf hybrid cryptosystems}, where an initial expensive private channel is established with public key to securely share a symmetric key, which will be used for lightweight communication.

Modern cryptography bases the identity of entities (e.g. banks) on their {\bf public keys}: a genuine entity can {\em mathematically} prove that they are the intended entity of the communication by demonstrating (without revealing) that they hold the {\bf private key} corresponding to the public one.

As the name suggests, public keys are not kept secret.
Once a message is encrypted with a public key, only the holder of the private key related to that public key can decrypt the cyphertext.
Conversely, encrypting a message with a private key will allow anybody to inspect the plaintext, along being (mathematically) certain of the authenticity of the message, as only the owner of the private key could have produced the encrypted message: this is the principle behind {\bf digital signatures}.
Digitally signing messages\footnote{In pair with the receiver's identity to avoid subtle attacks.} with the private key of the sender, then encrypting the resulting message with the public key of the receiver can offer both authenticity and confidentiality.

As the communication flows through insecure channels, there is no guarantee that the public key {\em received} from the channel is the right one in the first place, as it might have been tampered with by an attacker.
\begin{figure}[htbp]
  \caption{The basic structure of mutual authentication in a generic PKI.}
  \label{fig:pki-arch-easy}
  \begin{center}
    \includegraphics[width=.7\textwidth]{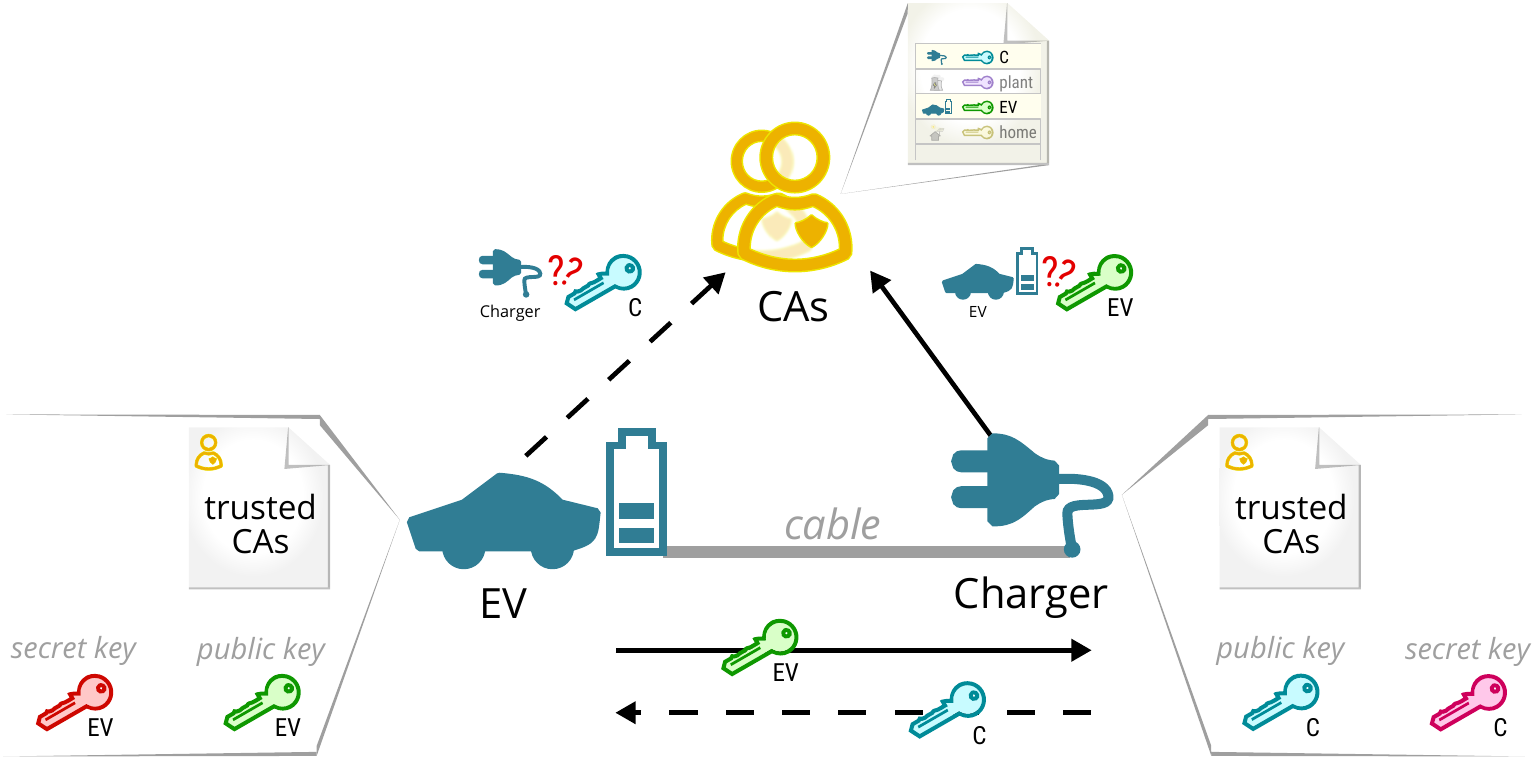}
  \end{center}
\end{figure}
One way to solve this is for devices to store a list of {\em trusted} entities, called {\bf Certificate Authorities} (CAs).
CAs store {\bf certificates}, which bind entities to unique {\bf public keys}\footnote{CAs do not hold other entities' private keys}.
Certificates are {\em securely} deposited into CAs, which are trusted servers.
So to identify themselves, entities (e.g. banks, websites) publish their certificates; any user can automatically verify the certificate against CAs, as shown in Figure~\ref{fig:pki-arch-easy} for the example of a connection between an EV and a charger.

CAs are part of a service/trust framework responsible for creating, revoking, distributing, and updating key pairs and certificates.
We denote any such framework as a Public-Key Infrastructure (PKI)\footnote{Not every PKI requires central trusted CAs (e.g. web of trust), but the CA-based PKI here described is the most common over the Internet. More details can be found in the popular textbook: Katz, Jonathan, and Yehuda Lindell. {\bf Introduction to modern cryptography}. CRC press, 2020.}.
In essence, the users do not need to hold all certificates, but only carry certificates belonging to CAs; then the users can query CAs for any secure connection they need to establish with authenticated entities.

Generally, the end-users are not known to CAs and need other explicit ways to be authenticated.
So for example, an account holder needs to use a password, a secret code, or a fingerprint to be authenticated by their bank, whereas the bank uses one certificate verifiable against the CA without manual intervention.

\well[technical note]{
  The security of the public key cryptography relies on the complexity for an adversarial entity to reconstruct the private key even knowing the corresponding public key.
Complexity guarantees that the amount of time that the adversary would need to crack the private key can be increased by increasing a security parameter, i.e. the bit length of the key.
Common security parameters, e.g. 2048-bit keys, are estimated to require several centuries to succeed for an attacker with all computing power in the world focused on cracking one key with known methodology.
}

\begin{figure}[htbp]
  \caption{Example of proposed PKI architecture \& governance model for an industry, such as the EV ecosystem. Adapted from: Mike Nelson, DigiCert \& Oscar Marcia, Eonti Inc. - Public Key Infrastructure (PKI) for electric vehicles, \url{https://youtu.be/nEBJzPVZNd0}.}
  \label{fig:pki-arch}
  \includegraphics[width=\textwidth]{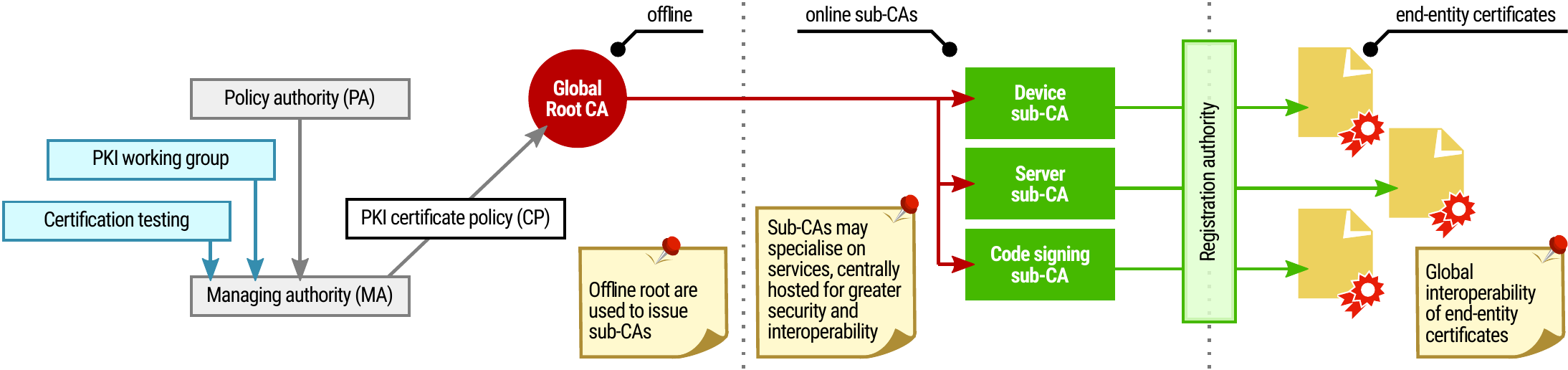}
\end{figure}

We briefly described keys and PKI, we now look at the EV ecosystem.
The main PKI deployment in the EV ecosystem is with the Plug\&Charge (PnC) seamless service described by ISO 15118.
Hubject has been operating the ISO15118 Plug\&Charge for some years but it has been announced (circa Q2 2021) that CharIN e.V. will be managing its PKI instead and aiming for a Europe-wide rollout\footnote{\url{https://www.charin.global/news/plug-and-charge-europe-enabled-by-charin-is-soon-to-be-rolled-out/}}.
Plug\&Charge aims for automated authentication, authorisation and billing of the charging event.
With the aid of a PKI, the only action required by the driver to charge their car is to connect the car to the charger and no additional authentication methods are needed (e.g. no RFID card, no signing through a phone App, etc.).
For the PnC service to work, the car (at the very least\footnote{ISO 15118 outlines a PKI where 5 entities are required to have a certificate.}) needs to have a registered certificate.
This is a challenge as the CAs are supposed to store a sensibly higher amount of certificates\footnote{This is comparable as if banks store a certificate for each {\em potential} customer.}.
Moreover, other desirable properties add up to the overall complexity, e.g. expiration, suspension and revocation of certificates, or trustability of CAs (transparency, security, availability and constant audits for compliance with security industry standards).

Designing a PKI for the current number of EVs is as easy as using the existing PKI over the Internet, i.e. registering a certificate for any entity that needs to be authenticated.
However, designing a PKI that scales (more than 1 billion cars are currently hitting the road) is non-trivial and needs to take into account the different context and requirements of the EV charging infrastructure: the high mobility of vehicles, diversity of charging protocols and outlets, compliance to local and international restrictions and policies, and the high number of overall certificates to be stored (to have seamless functionality, as PnC).
A recent project assessed the PKI described in ISO 15118 and how it could be improved~\cite{chargepoint2019practical}.
Moreover, a dedicated research project by SAE is on-going\footnote{\url{ https://www.sae.org/news/press-room/2021/02/sae-international-hires-world-class-contractor-team-for-ev-charging-public-key-infrastructure-cooperative-research-project}}.
SAE is working with mobility and security experts to drive the creation and operation of a common worldwide EV charging industry PKI.
It is worth noting (circa 2020) that core team members of the SAE project don't include grid operators. 
Other PKI work for EV charging is done by ElaadNL who proposed PKI designs compatible with ISO 15118 and coexisting with a peer-to-peer PKI~\cite{elaadnl2018exploring}.

Figure~\ref{fig:pki-arch} illustrates an example of a proposed PKI architecture and governance model for an industry, which can be used for the EV ecosystem.
This is a simple 2-tier architecture with one offline root CA as the trust anchor but, if implemented correctly, could provide extensive security measures for a complex industry involving several entities such as the EV ecosystem.
What ensures the success of a PKI is a good governance model behind the infrastructure and technology.
The Policy Authority (PA) and the Managing Authority (MA) provide the governance and policy framework for this system.
Borrowing an example from the aviation industry (e.g. AeroMACS), the PA acts as an advisory council to the MA, and it is made up of aviation administrations from several regions (i.e. Eurocontrol, Civil Aviation Administration of China and the US Federal Aviation Administration).
In addition to taking input from the PA, there are several working groups that inform the work of the MA.
These working groups come together and look at technical requirement of the system.
There is also certification testing for devices to ensure they are compliant to the agreed standards.
All of this feeds into the certificate policy that governs the PKI~\cite{nelson2020public}, see Figure~\ref{fig:pki-arch}. 

The design of a PKI for EV charging infrastructure, including smart charging and V2G, can build on top of the existing technology used for the secure functioning of the Internet and other domains (e.g. banking industry, airport operations).
It should also build on top of on-going PKI work for EV charging infrastructure.
However, additional work is required to ensure the PKI is protocol neutral, extensible, scalable, and crucially enabling for smart charging and V2G use cases. 

We envision that mobility and energy industries in the UK would collaborate together to design a PKI dedicated for EV charging.
Crucially, it is key that UK stakeholders participate in on-going European and International efforts on the topic, as to the best of our knowledge we have not noticed their participation.  
\small
\renewcommand{\bibname}{References}
\printbibliography

\newpage
\thispagestyle{empty}
\begin{tikzpicture}[remember picture,overlay,background rectangle/.style={fill=palette55!70!white}, show background rectangle]
  \node[] (frontpage) at (current page.center) { \includegraphics[width=\paperwidth]{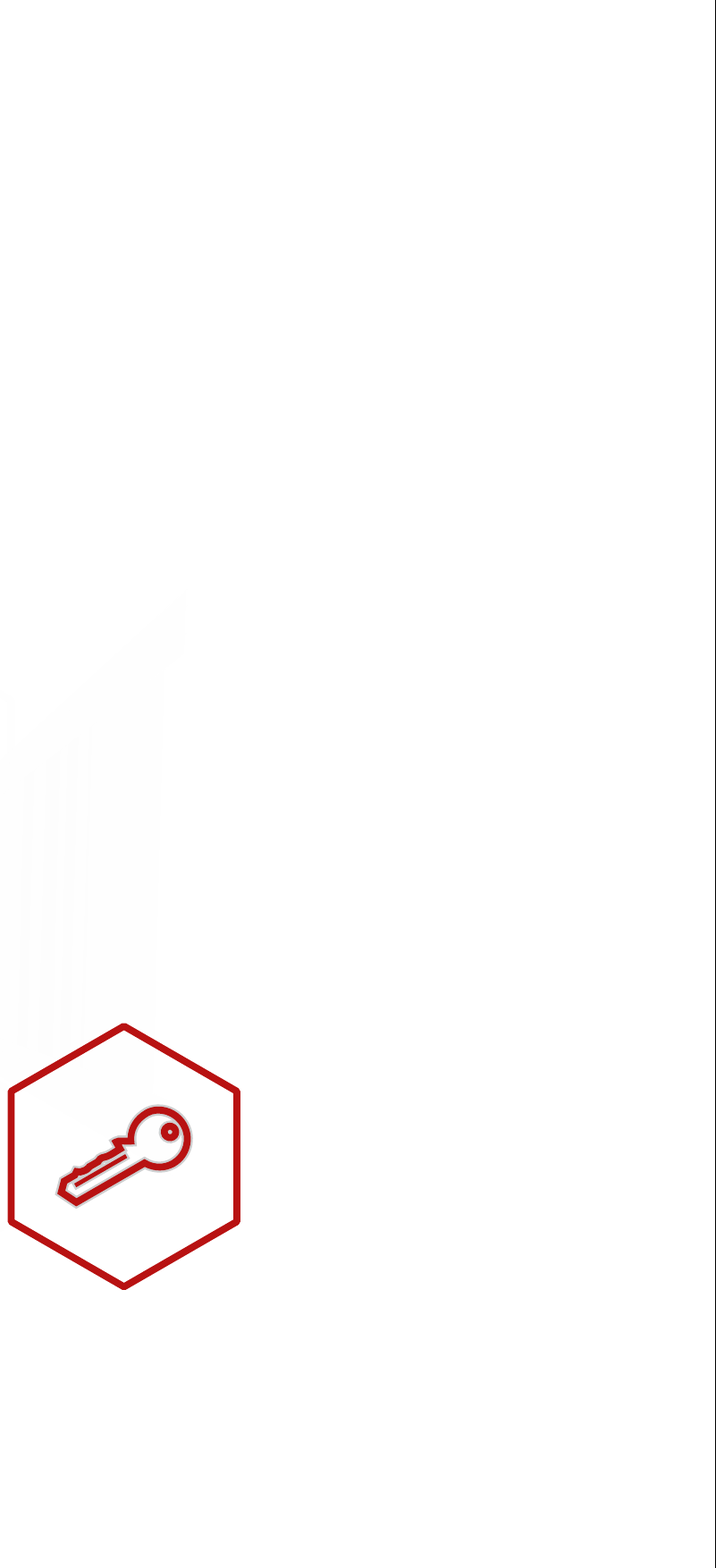} };
  
  \node[anchor=north west,xshift=3em,yshift=-8em] (names) at (current page.north west) {
    {\color{titles}\bfseries {\textsc{Roberto Metere\quad Myriam Neaimeh\quad Charles Morisset}}}
  };
  \node[anchor=north west,xshift=3em,yshift=-9em] (names) at (current page.north west) {
    {\color{titles}\bfseries {\textsc{Carsten Maple\quad Xavier Bellekens\quad Ricardo M. Czekster}}}
  };
  \node[anchor=north west,xshift=3em,yshift=-13em] (names) at (current page.north west) {
    \begin{varwidth}{0.5\paperwidth}
      {\color{subtitles}\Large\bfseries \nohyphens{\textsc{\inserttitle}}}
    \end{varwidth}
  };
  
  \node[anchor=north west,minimum height=4cm] (names) at ([xshift=1em,yshift=4.2cm]current page.south west) {
    \begin{tabularx}{0.7\paperwidth}{cccc}
        \includegraphics[height=1.2cm,valign=b]{res/ncl-logo.pdf}
      & \includegraphics[height=1.2cm,valign=b]{res/ati-logo.pdf}
      & \includegraphics[height=1.0cm,valign=b]{res/warwick.pdf}
      & \includegraphics[height=1.8cm,valign=b]{res/strath.pdf}
    \end{tabularx}
    };

\end{tikzpicture}
\end{document}